\documentclass{vldb_response}

\usepackage{etoolbox}
\usepackage[noend]{algpseudocode}
\usepackage{algorithm,algorithmicx}
\usepackage{graphicx}
\usepackage{listings}
\usepackage{url}
\usepackage{color}
\usepackage{xcolor}
\usepackage{enumerate}
\usepackage{lipsum} 
\usepackage{txfonts}
\usepackage{pbox}
\usepackage{wasysym}
\usepackage{amssymb}
\usepackage{subfig}
\usepackage{float}
\usepackage{paralist}
\usepackage{enumitem}
\usepackage{balance}
\let\thepage\relax

\newcommand{\specialfigure}[0]{
\begin{figure*} [t!]
\centering
    \includegraphics[scale=0.25]{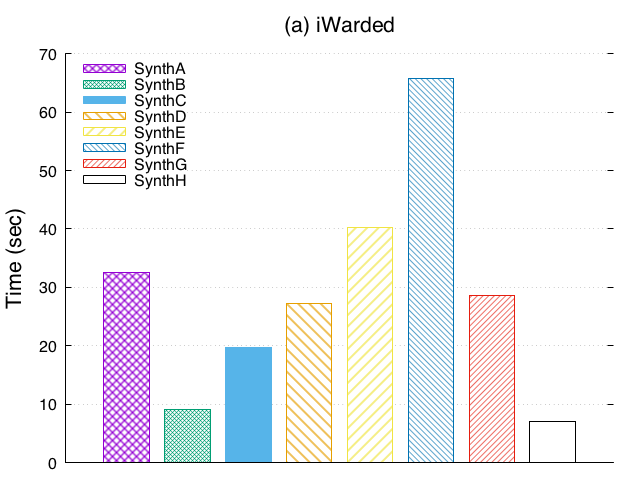}
    \includegraphics[scale=0.25]{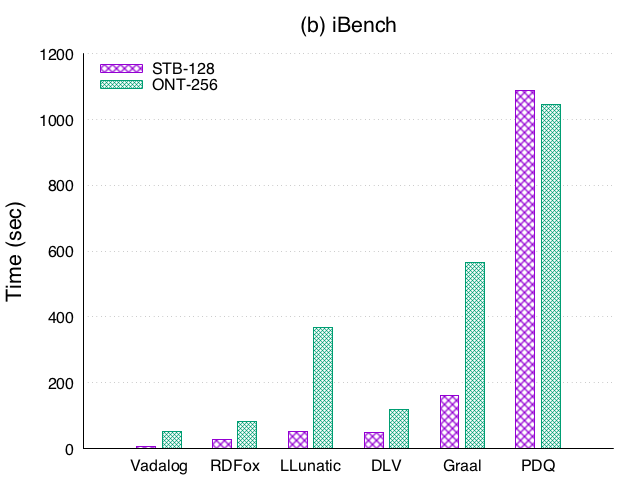}
    \includegraphics[scale=0.25]{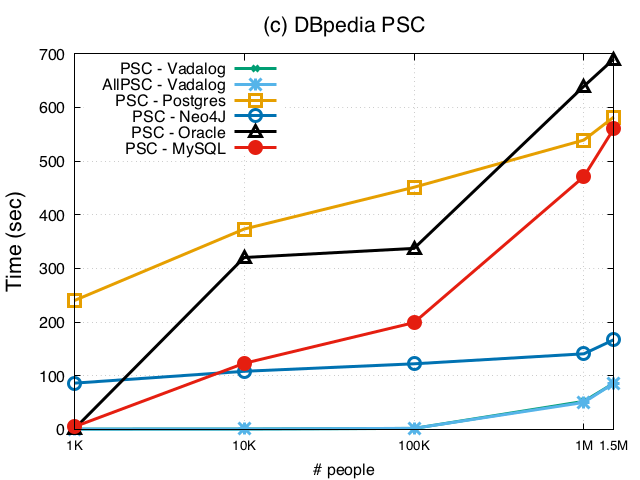}\\
    \includegraphics[scale=0.25]{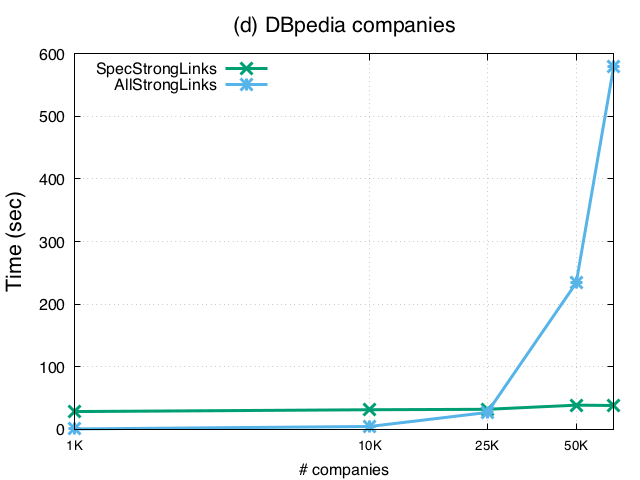}
    \includegraphics[scale=0.25]{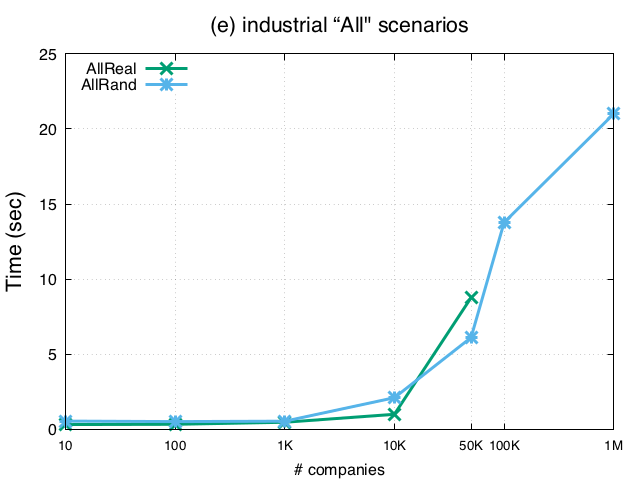}
    \includegraphics[scale=0.25]{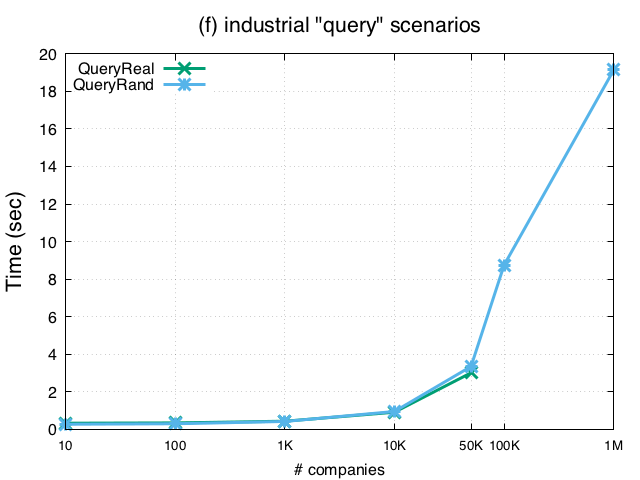}\\
    \includegraphics[scale=0.25]{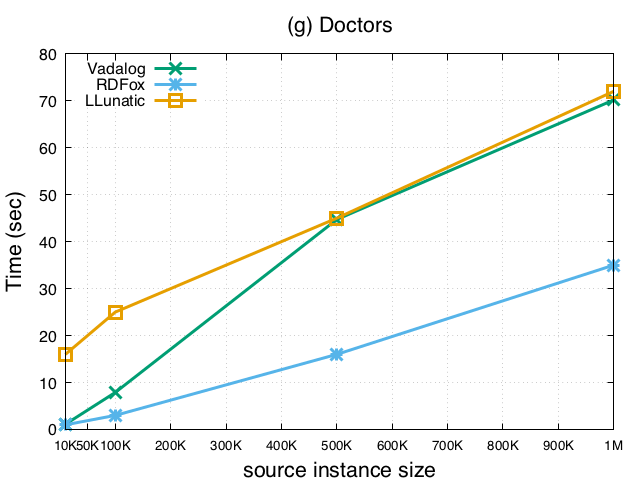}
    \includegraphics[scale=0.25]{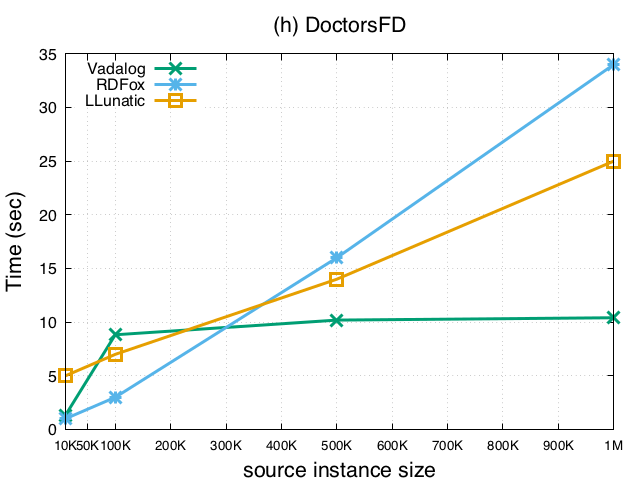}
    \includegraphics[scale=0.25]{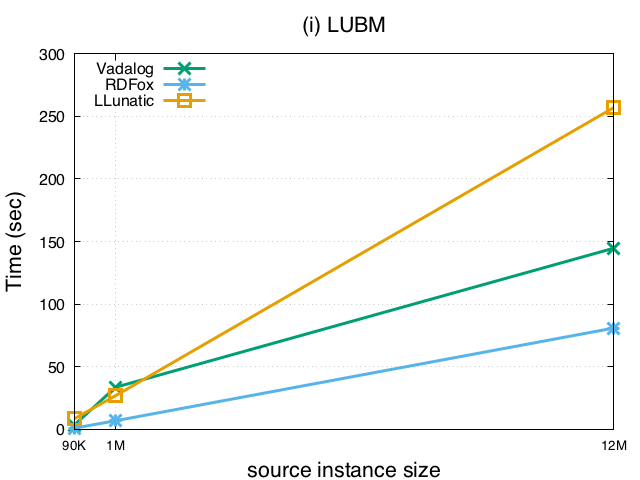}
\caption{Reasoning times for the experimental scenarios.}
\label{fig:experiments}
\end{figure*}
}

\newcommand{\cbstart}[0]{}
\newcommand{\cbend}[0]{}
\newcommand{\specialbreak}[0]{}
\newcommand{\figurehere}[0]{\specialfigure}

\vldbTitle{The Vadalog System: Datalog-based Reasoning for Knowledge Graphs}
\vldbAuthors{Luigi Bellomarini, Emanuel Sallinger, and Georg Gottlob}
\vldbVolume{11}
\vldbNumber{9}
\vldbYear{2018}
\vldbDOI{https://doi.org/10.14778/3213880.3213888}

\begin{document}

\definecolor{darkgreen}{rgb}{0.0,0.6,0.00}

\newcommand{\LBComments}[1]{\textcolor[rgb]{0.60,0.6,0.00}{[LB:] #1}} 
\newcommand{\PAComments}[1]{\textbf{\textcolor[rgb]{1.00,0.00,0.00}{[PA:] #1}} }
\newcommand{\shortpaper}[1]{#1}
\newcommand{\longpaper}[1]{#1}

\newcommand{\new}[1]{\textcolor[rgb]{0.00,0.0,0.00}{#1}}
\newcommand{\newspecial}[1]{\textcolor[rgb]{0.00,0.0,0.00}{#1}}
\newcommand{\old}[1]{}

\newcommand{\movedfrom}[1]{}
\newcommand{\movedto}[1]{\textcolor[rgb]{1.00,0.00,0.00}{#1}}

\newcommand{\shorten}[1]{#1}

\newcommand{\shortAdd}[1]{#1}
\newcommand{\shortRemove}[1]{#1}
\newcommand{\vsRemove}[1]{#1}
\newcommand{\vsAdd}[1]{#1}

\newtheorem{example}{Example}
\newtheorem{definition}{Definition}
\newtheorem{theorem}{Theorem}

\title{The Vadalog System:\\ Datalog-based Reasoning for Knowledge Graphs}
\numberofauthors{3}
\author{
\alignauthor
Luigi Bellomarini\\
       \affaddr{University of Oxford and\\ Banca d'Italia and\\ Universit\`a Roma Tre}
\alignauthor
Emanuel Sallinger\\
       \affaddr{University of Oxford}
\alignauthor Georg Gottlob\\
       \affaddr{University of Oxford and\\ TU Wien}
}

\maketitle

\newcommand{\originalAbstract}{
Over the past years, there has been a resurgence of Datalog-based systems in the database community as well as in industry. In this context, it has been recognized that to handle the complex knowl\-edge-based scenarios encountered today, such as reasoning over large knowledge graphs, Datalog has to be extended with features such as existential quantification. Yet, Datalog-based reasoning in the presence of existential quantification is in general undecidable. Many efforts have been made to define decidable fragments. Warded Datalog+/- is a very promising one, as it captures the PTIME complexity class while allowing ontological reasoning. Yet so far, no implementation of Warded Datalog+/- was available.

In this paper we present the Vadalog system, a Datalog-based system for performing complex logic reasoning tasks, such as those required in advanced knowledge graphs. The Vadalog system is Oxford's contribution to the VADA research programme, a joint effort of the universities of Oxford, Manchester and Edinburgh and around 20 industrial partners.
As the main contribution of this paper, we illustrate the first implementation of Warded Datalog+/-, a high-performance Datalog+/- system utilizing: an aggressive recursion and termination control strategy leveraging the properties of  Warded Datalog+/-; a modular architecture that makes the most of the available experience in the development of database systems; an in-memory stream-based architecture, which combines a query-driven approach with a pervasive use of local caching. We conclude with a comprehensive experimental evaluation of the system. 
}

\section*{ABSTRACT}
Over the past years, there has been a resurgence of Datalog-based systems in the database community as well as in industry. In this context, it has been recognized that to handle the complex knowl\-edge-based scenarios encountered today, such as reasoning over large knowledge graphs, Datalog has to be extended with features such as existential quantification. Yet, Datalog-based reasoning in the presence of existential quantification is in general undecidable. Many efforts have been made to define decidable fragments. Warded Datalog+/- is a very promising one, as it captures PTIME complexity while allowing ontological reasoning. Yet so far, no implementation of Warded Datalog+/- was available.
In this paper we present the Vadalog system, a Datalog-based system for performing complex logic reasoning tasks, such as those required in advanced knowledge graphs.
The Vadalog system is Oxford's contribution to the VADA research programme, a joint effort of the universities of Oxford, Manchester and Edinburgh and around 20 industrial partners.
As the main contribution of this paper, we illustrate the first implementation of Warded Datalog+/-, a high-performance Datalog+/- system utilizing an aggressive termination control strategy. 
We also provide a comprehensive experimental evaluation.

\section{introduction}
\label{sec:introduction}

\noindent
The importance of capitalizing and exploiting corporate knowledge has been 
clear to decision makers since the late 1970s, when this idea was gradually
made concrete in the context of \emph{expert systems}, software frameworks
able to harness such knowledge and provide answers to structured business questions.
Through \emph{deductive database systems} --  database systems that have advanced reasoning capabilities -- and in particular the language \emph{Datalog}, the area became of high interest to the database community in the 1980s and 1990s. Yet, even though the importance of harnessing knowledge certainly has grown steadily since then, culminating in today's desire of companies to build and exploit \emph{knowledge graphs}, the interest in deductive databases has faltered.

Many factors have prevented an effective large-scale development
and diffusion of deductive database systems:  immature hardware technology,
unable to cope with the rising challenges and opportunities coming from the
Big Data applications; the rigidity of existing database management systems, unable to go beyond the standard requirements of query answering; the lack of knowledge languages expressive enough to address real-world cases.

The situation has rapidly changed in recent times, thanks to accelerations in many fields.
The strong need for scalability firstly represented by Internet giants and promptly
understood by large and medium-sized companies led to the soar of a flurry of
specialized middleware modules having scalability as a central target. 
The resurgence of \emph{Datalog} in academia and industry \cite{ACGK15,conf/datalog/2012,CaGK13,CaGL12,CGLMP10,CaGP12} turned out to be a key factor. Companies like LogicBlox  have proven that a fruitful exchange between academic research and high-performance industrial applications can be achieved based on Datalog \cite{ACGK15}, and companies like LinkedIn have shown that the interest in Datalog permeates industry \cite{MoustafaPYD16}.
Meanwhile, the recent interest in machine learning
brought renewed visibility to AI, raising interest and triggering investments in thousands of companies
world-wide, which suddenly wish to collect, encapsulate and exploit their corporate knowledge \vsAdd{in the form of a knowledge graph}.

\cbstart
\new{The term \emph{knowledge graph} (KG) has no standard definition. It can be seen as referring only to Google's Knowledge Graph, to triple-based models, or to multi-attributed graphs, which represent $n$-ary relations \cite{0001KT17,UrbaniJK16}. As shown by Kr{\"o}tzsch~\cite{Krotzsch11}, in order to support rule-based reasoning on such data structures, 
it is sometimes necessary to use tuples of arity higher than three at least for intermediate results.
In this paper, we adopt a general notion of KGs by allowing relations of arbitrary arity, to support all of these models and modes of reasoning.}
\cbend

\cbstart
\begin{example}
\label{ex:kroetzsch}
\new{An  example of a simple knowledge graph reasoning setting is given in~\cite{0001KT17}:}

\noindent
\begin{center}
\new{
${\rm Spouse}(x,y,start,loc,end) \rightarrow {\rm Spouse}(y,x,start,loc,end)$}
\end{center}

\noindent
\new{This rule expresses that when a person $x$ is married to a person $y$ at a particular location, starting date and end date, then the same holds for $y$ and $x$. That is, the graph of persons and their marriage relations is symmetric.
}
\end{example}

\noindent
\new{As stated in~\cite{0001KT17}, most modern ontology languages are not able to express this example.}
\new{
Beyond this simple example, there are numerous requirements for a system that allows ontological reasoning over KGs.
Navigating graphs is impossible without powerful recursion;} ontological reasoning is impossible without existential quantification in rule heads \cite{GoPi15}.\vsRemove{Yet reasoning with recursive Datalog is undecidable in the presence of existential quantification, so some tradeoffs have to be accepted.} 
\new{An analysis of various requirements was given in~\cite{BGPS17}.
In this paper, we isolate three concrete requirements for reasoning over knowledge graphs:
}

\begin{enumerate}[leftmargin=4mm]
\addtolength{\itemsep}{-.5em}
\item \new{\textbf{Recursion over KGs}. Should be at least able to express full recursion and joins, i.e., should at least encompass Datalog.
Full recursion in combination with arbitrary joins allows to express complex reasoning tasks over KGs. Navigational capabilities, empowered by recursion, are vital for graph-based structures.
}
\item \new{\textbf{Ontological Reasoning over KGs}. Should at least be able to express SPARQL reasoning under the OWL 2 QL entailment regime and set semantics.
OWL~2~QL is one of the most adopted profiles of the Web Ontology Language, standardized by W3C.
}
\item \new{\textbf{Low Complexity}. Reasoning should be tractable in data complexity. Tractable data complexity is a minimal requirement for allowing scalability over large volumes of data.
}
\end{enumerate}

\noindent
\new{Beyond these specific requirements, the competition between powerful recursion, powerful existential quantification and low complexity} \cbend has spawned fruitful research throughout the community as reasoning with recursive Datalog is undecidable in the presence of existential quantification.
This has been done under a number of different names, but which we shall here call \emph{Datalog$^\pm$}, the ``+'' referring to the additional features \new{(including existential quantification)}, the ``-'' to restrictions that have to be made to obtain decidability. Many languages within the Datalog$^\pm$ family of languages have been proposed and intensively investigated~\cite{BagetLM10,BagetMRT11,CaGK13,CaGL12,CGLMP10,CaGP12}. \cbstart \new{Details on the languages are given in Section~\ref{sec:language}.  Depending on the syntactic restrictions, they achieve a different balance between expressiveness and computational complexity.}

\new{
Figure 1 gives an overview of the main Datalog$^\pm$ languages. 
In fact, most of these candidates, including Linear Datalog$^\pm$, Guarded Datalog$^\pm$, Sticky Datalog$^\pm$ and Weakly Sticky Datalog$^\pm$ do not fulfil (1). 
Datalog itself does not fulfil (2). Warded and Weakly Frontier Guarded Datalog$^\pm$ satisfy (1) and (2), thus are expressive enough. However, the expressiveness of Weakly Frontier Guarded Datalog$^\pm$ comes at the price of it being EXPTIME-complete~\cite{BagetMRT11}. Thus it does not fulfil (3).
} 

\new{Thus, in total, the only known language that satisfies (1), (2) and (3) is Warded Datalog$^\pm$.} \cbend Yet, while Warded Datalog$^\pm$ has very good theoretical properties, the algorithms presented in \cite{GoPi15} are alternating-time Turing machine algorithms, far away from a practical implementation.

\cbstart
\begin{figure}[h!]
\centering
    \includegraphics[scale=0.38]{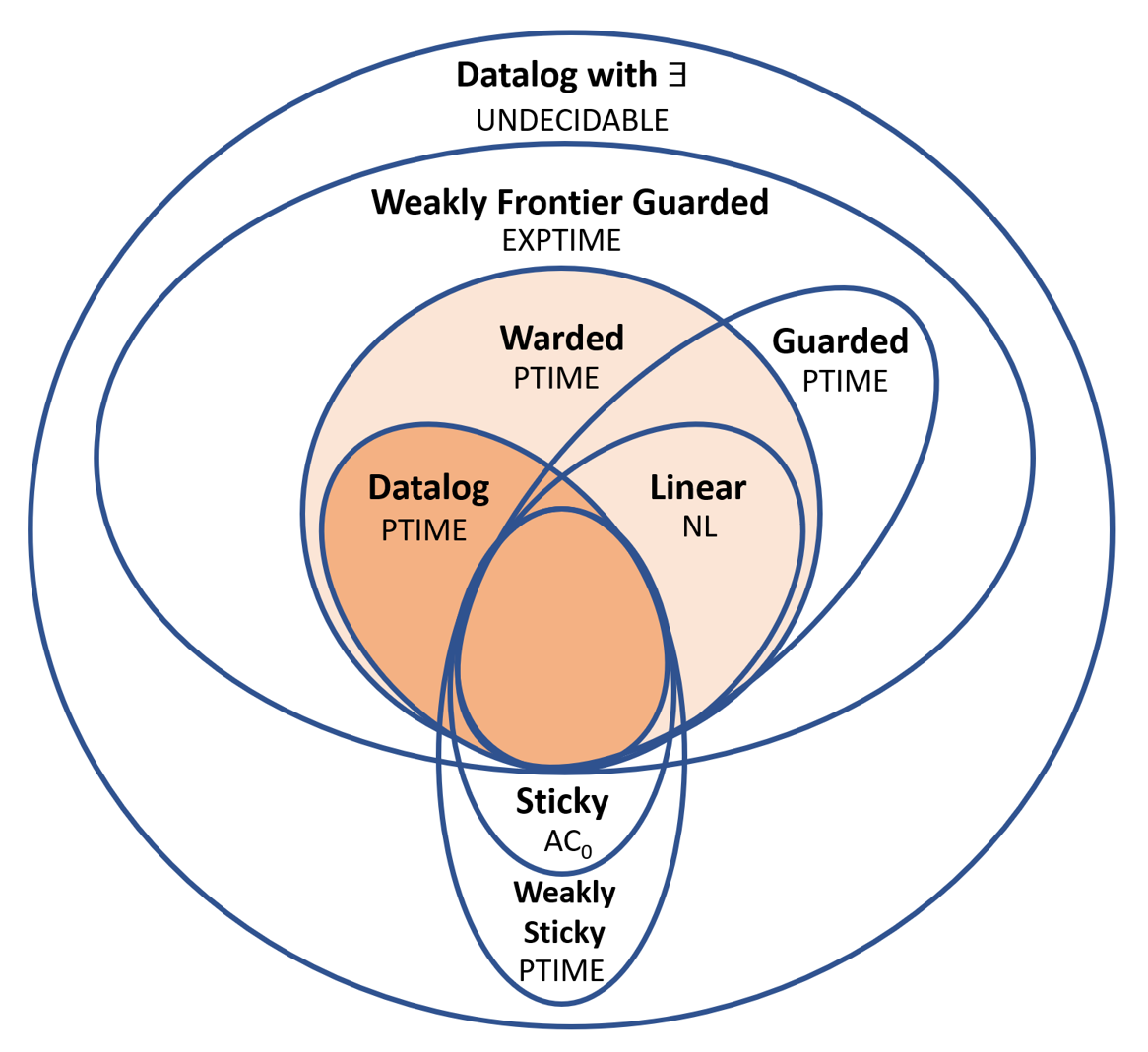}
\caption{\newspecial{Syntactic containment of Datalog$^\pm$ languages. Annotations (non-bold) denote data complexity. All names that do not explicitly mention Datalog refer to the respective Datalog$^\pm$ languages. E.g., ``Sticky'' refers to ``Sticky Datalog$^\pm$''.}}
\label{fig:expressive}
\end{figure}
\cbend

\old{
\medskip
\noindent
Recently, a vision and reference architecture for \textbf{knowledge graph management systems}
(KGMS)~\cite{BGPS17} was presented, bringing together the traditional expertise of the database community, the desire of companies to build knowledge graphs, and the larger interest of the analytics and AI community.
The vision calls for a Datalog-based deductive database system at the core, and identifies the language called Warded Datalog$^\pm$ as a very good candidate for its logical foundation.
}
\shortRemove{
Before going into the details, let us consider an example of a Datalog program that is at the same time close to a classical example \cite{datalog1} as well as close to an application of the knowledge graph of one of our current industrial partners. Note that the \texttt{msum} operator, i.e., aggregation in the form of \emph{monotonic sum}, used in the following program is not a standard feature of Datalog, but an extension that some Datalog-based systems support. It will be discussed later.

\begin{example}
\label{ex:running}
Consider a set of rules about ownership relationships among a large number of companies.  The extensional knowledge is stored in a database in the form of tuples of a relation ${\rm Own}(\mathit{comp_1,\ comp_2},w)$: company $\mathit{comp_1}$ directly owns a fraction $w$ of company $\mathit{comp_2}$, with $0 \leq w\leq 1$.
In addition, with a set of two rules, we intensionally represent the  concept of \emph{company control} (also studied in~\cite{datalog1}) as encoded by the set of rules in Example~\ref{ex:running}: 
a company $x$ controls a company $y$ if company $x$ directly owns more than half of the shares of company $y$ or if $x$ controls a set $S$ of companies that jointly own more than half of $y$.

\begin{eqnarray*}
{\rm Own}(x,y,w), w > 0.5 &\rightarrow& {\rm Control}(x,y)\\
{\rm Control}(x,y), {\rm Own}(y,z,w),&&\\
v=\texttt{msum}(w,\langle y  \rangle), v>0.5 &\rightarrow& {\rm Control}(x,z).
\end{eqnarray*}

\noindent
Here, for fixed $x$, the aggregate construct $\texttt{msum}(w, \langle y \rangle)$ forms the sum over all values $w$ such that for some company $y$, ${\rm Control}(x,y)$ is true, and ${\rm Own}(y,z,w)$
holds, i.e., company $y$ directly owns fraction $w$ of company $z$.
Now, with the described knowledge graph, many questions can be asked, such as:
1. obtain all the pairs \((x,y)\) such that company \(x\) controls company \(y\);
2. which companies are controlled by \(x\)? Which companies control \(x\)? 3. does company \(x\) control company \(y\)?
\end{example}
}
\old{
\noindent
Yet, during the resurgence of Datalog, it has been recognized that to handle the complex knowledge-based scenarios encountered today, Datalog has to be extended with features such as existential quantification. Ontological reasoning on knowledge graphs is impossible without existential quantification in rule heads \cite{GoPi15}.
}
\old{
However important, the great drawback is that Datalog-based reasoning in the presence of existential quantification is undecidable in general. This spawned fruitful research throughout the community, under a number of different names, but which we shall here call \emph{Datalog$^\pm$}, the ``+'' referring to the additional features, the ``-'' to restrictions that have to be made to obtain decidability. Many languages within the Datalog$^\pm$ family of languages have been proposed and intensively investigated \cite{BagetLM10,BagetMRT11,CaGK13,CaGL09,CGLMP10,CaGP12}. Warded Datalog$^\pm$ \cite{GoPi15} was recognized to be an extremely strong candidate for the requirements of \cite{BGPS17}.}\old{
Warded Datalog$^\pm$ is a language that
\begin{itemize}[leftmargin=*]
\item includes the entirety of Datalog with stratified negation, understood by academic and practitioners' communities, with clear semantics.
\item includes existential quantification with mild restrictions \cite{GoPi15}.
\item is able to express SPARQL queries under set semantics \cite{PerezAG09} and the entailment regime for
OWL 2 QL ~\cite{GoPi15}, and is able to perform ontological reasoning.\footnote{
A very interesting example covered by the language is given in \cite[Section 4]{Krotzsch17}, a paper that discusses ontologies and knowledge graphs. We cannot present the example here for space reasons.}
\end{itemize}
}

\medskip 
\noindent
In this paper we present \textbf{the Vadalog system}, Oxford's contribution
to the {\sc VADA} research project~\cite{vadawebpage,KonstantinouKAC17}, a joint effort of the universities
of Oxford, Manchester and Edinburgh and around 20 industrial partners such as Facebook, BP, and the NHS (UK national health system). The Vadalog system is built around the {\sc Vadalog} language, with Warded Datalog$^\pm$ as its logical core.
It is currently used as the core deductive database system of the overall Vadalog Knowledge Graph Management System described in \cite{BGPS17} as well as at various industrial partners, including the finance, security, and media intelligence industries.
The \textbf{main contributions} of this paper are:

\begin{itemize}[leftmargin=*]
\item A \textbf{novel analysis} of Warded Datalog$^\pm$, focused on practical implementation. In particular, we identify a number of \emph{guide structures} that closely exploit the underpinnings of Warded Datalog\(^\pm\) to guarantee termination while keeping the size of such guide structures limited to ensure a small memory footprint. These three guide structures, the \emph{linear forest}, \emph{warded forest}, and \emph{lifted linear forest} are related, but play complementary roles for exploiting the periodicity of execution. The culmination of this analysis is the \textbf{first practical algorithm for Warded Datalog$^\pm$}.

\item A \textbf{system and architecture} that implements this algorithm in a relational database-inspired operator pipeline architecture. 
The pipeline's operators rely on \textit{termination strategy wrappers} which transparently prevent the generation of facts that may lead to non-termination while ensuring the correctness of the result. 
\shortRemove{The system is completed by a wide range of standard and novel optimization techniques such as the dynamic construction of in-memory indices, stream-optimized ``slot machine'' joins, and mo\-notonic aggregation.}

\item A full-scale \textbf{experimental evaluation} of the 
Vadalog system on a variety of real-world and synthetic scenarios that thoroughly validate
the effectiveness of our techniques on Warded Datalog$^\pm$ in absolute terms and
comparatively with the top existing systems, which are outperformed by our reasoner.
\end{itemize}

\medskip
\noindent
\textbf{Overview}.
In Section~\ref{sec:language} we describe the {\sc Vadalog} language. 
In Section~\ref{sec:termination} we illustrate how we use the properties
of Warded Datalog$^\pm$ to handle recursion and termination.
In Section~\ref{sec:architecture} we present the architecture of the Vadalog
system. Section~\ref{sec:experiments} is dedicated to a full-scale evaluation
of the system. In Section~\ref{sec:relwork} we discuss the related work \new{and compare with other systems}. The conclusions
are in Section~\ref{sec:conclusions}.

\section{Reasoning with VADALOG}
\label{sec:language}

\noindent
Our system adopts {\sc Vadalog} as a reference language. 
Its core is based on Datalog\(^\pm\) and in particular on Warded Datalog\(^\pm\). 
Before discussing the details of Warded Datalog\(^\pm\) and the specific extensions
to it that we have in {\sc Vadalog}, we first 
 recall some foundations underlying Datalog\(^\pm\).

\vsRemove{\subsection{The Core Language}}
\label{sec:datalog}

Let $\mathbf{C}$, $\mathbf{N}$, and $\mathbf{V}$ be disjoint countably infinite sets of {\em constants}, {\em (labeled) nulls} and (regular) {\em variables}, respectively. A {\em (relational) schema} $\mathbf{S}$ is a finite set of relation symbols (or predicates) with associated arity. A {\em term} is a either a constant or variable. An {\em atom} over $\mathbf{S}$ is an expression of the form $R(\bar v)$, where $R \in \mathbf{S}$ is of arity $n > 0$ and $\bar v$ is an $n$-tuple of terms. 
A {\em database instance} (or simply \textit{database}) over $\mathbf{S}$ associates to each relation symbol in $\mathbf{S}$ a relation of the respective arity over the domain of constants and nulls. The members of relations are called \textit{tuples}. By some abuse of notations, we sometimes use the terms atom, tuple and fact interchangeably.

Datalog\(^\pm\) languages consists of \emph{existential rules}, or
\emph{tuple-genera\-ting dependencies}, which generalize Datalog rules with existential
quantification in rule heads.
Such a rule is a first-order sentence of the form
\(
\forall \bar x \forall \bar y (\varphi(\bar x,\bar y)\ \rightarrow\ \exists \bar z \, \psi(\bar x, \bar z))
\),
where $\varphi$ (the {\em body}) and $\psi$ (the {\em head}) are conjunctions of atoms with constants and variables.
For brevity, we write this existential rule as $\varphi(\bar x,\bar y) \rightarrow \exists \bar z\, \psi(\bar x,\bar z)$ and replace $\wedge$ with comma to denote conjunction of atoms.  If there is at most one atom in $\varphi$, we call a rule \textit{linear}, otherwise \textit{non-linear}.

The intuitive meaning of such a rule is as follows: if there is a fact $\varphi(\bar t,\bar t')$ that occurs in an instance $I$, then there exists a tuple ${\bar t}''$ of constants and nulls such that the facts $\psi(\bar t,\bar t'')$ are also in $I$.
Formally, the semantics of a set of existential rules $\Sigma$ over a database $D$, denoted $\Sigma(D)$, is defined via the well-known \emph{chase procedure}. Roughly, the chase adds new facts to $D$ (possibly involving null values used to satisfy the existentially quantified variables) until the final result $\Sigma(D)$ satisfies all the existential rules of $\Sigma$. Notice that, in general, $\Sigma(D)$ is infinite.  Example~\ref{ex:company_and_kp} is a simple such example:

\begin{example}
\label{ex:company_and_kp}
${\rm Company}(x) \rightarrow \exists p~{\rm {\underline{KeyPerson}}(\hat p,x)}$\\
${\rm Control}(x,y), { {\rm KeyPerson}}(\hat p,x) \rightarrow {\rm {\underline{KeyPerson}}(\hat p,y)}.$

\smallskip \noindent
The first rule
expresses that for every company \(x\) there exists a key person \(p\). 
Then, if company \(x\) controls company \(y\) and \(p\) is a key person for \(x\),
then \(p\) will be a key person for \(y\) as well.
\end{example}

\noindent
Now, given a pair $Q = (\Sigma,{\rm Ans})$, where $\Sigma$ is a set of existential rules and ${\rm Ans}$ an $n$-ary predicate, 
the evaluation of a query $Q$ over a database $D$, 
denoted $Q(D)$, is defined as the set of tuples $Q(D) = \{{\bar t} \in \mathit{dom}(D)^n \mid {\rm Ans}(\bar t) \in \Sigma(D)\}$. 
\vsRemove{This notion can be easily generalized in the presence of a 
set {\bf Ans} of \(n\)-ary predicates so 
that $Q(D) =  \{{\bar t} \in \mathit{dom}(D)^n \mid {\rm Ans}_1(\bar t)\ldots,{\rm Ans}_n(\bar t) \in \Sigma(D)\}$ for ${\rm Ans}_1,\ldots,{\rm Ans}_n \in$ {\bf Ans}.}
Observe that in Example~\ref{ex:company_and_kp}, the predicate {\rm KeyPerson} has been underlined
to mean it is in $\rm{Ans}$.

Based on this setting, we are interested in the {\em universal tuple inference} 
reasoning task (which we will simply call \emph{reasoning task} for the sake of simplicity): given
a database \(D\) and a pair $Q = (\Sigma,{\rm Ans})$, find an instance \(J\), such that
a tuple ${\bar t} \in J$ if and only if ${\bar t} \in Q(D)$
and for every other instance \(J^\prime\) such that ${\bar t} \in J^\prime$ if and only if ${\bar t} \in Q(D)$, 
there is a homomorphism \(h\) from \(J\) to \(J^\prime\). In other terms, we are interested in finding the most
general answer to the reasoning task, which amounts to saying that such answer is homomorphic to
any other answer (henceforth we call it \emph{universal}).

\vsRemove{With reference to Example~\ref{ex:company_and_kp}, let us suppose
we have the following database instance:
\begin{eqnarray*}
D = \{{\rm Company}(a), {\rm Company}(b), {\rm Company}(c), \\
{\rm Control}(a,b), {\rm KeyPerson}(a,Bob), {\rm Control}(a,c)\}
\end{eqnarray*}
The following instances are solutions to the reasoning task.
\begin{eqnarray*}
J_1 = D \cup \{{\rm KeyPerson}(b,Bob), {\rm KeyPerson}(c,Bob),{\rm KeyPerson}(c,\nu_1)\}\\
J_2 = D \cup \{{\rm KeyPerson}(b,Bob), {\rm KeyPerson}(c,Bob)\}
\end{eqnarray*}
\noindent
In general, the problem of finding this answer is computationally very hard, in fact 
undecidable, even when $Q$ is fixed and only $D$ is given as input~\cite{CaGK13}.
Many restrictions of existential rules have been identified, which make the
above problem decidable. Each of these restrictions gives rise to a new Datalog\(^\pm\) language.}

\vspace{-.25em} \medskip \noindent
{\bf Warded Datalog\(^\pm\).} The core of the {\sc vadalog} language is Warded Datalog\(^\pm\), a recent member of the
Datalog\(^\pm\) family. 
\shortRemove{Its importance has been widely motivated and is sustained
by several favourable properties. It is data tractable, as it captures the {\textsf{PTIME} }complexity class. 
In particular, the universal tuple inference reasoning task is polynomial time, which makes the fragment
ideal for the core language of a reasoner; it captures Datalog, since, as we will see, any set of Datalog rules is
warded by definition; it is suitable for ontological reasoning, as it generalizes
ontology languages such as the OWL 2 QL profile of OWL;
it is suitable for querying RDF graphs and, actually TriQ-Lite 1.0~\cite{GoPi15}, is a Warded-based
language (with stratified and grounded negation) able to express SPARQL queries under set semantics and the
entailment regime for OWL 2 QL.}
The full {\sc vadalog} language is then obtained as an extension to the Warded Datalog\(^\pm\) core
with features of practical utility\longpaper{, which will be introduced in Section~\ref{sec:extensions}}.

The idea behind the notion of wardedness is taming the propagation
of nulls during the construction of the answer in the chase procedure.
\longpaper{This is done by 1. marking as ``dangerous'', the body variables that can bind to labelled nulls and
also appear in the heads (hence propagate the nulls); 2. constraining that
in each rule all such variables appear within one single body atom, the \emph{ward}.
The ward must be such that it shares with other atoms in the body only
``harmless'' variables, i.e., they can bind only to ground values in the chase.
An example of warded program follows.

\begin{example}
\label{ex:simple_warded}
\begin{eqnarray*}
{\rm P}(x) \rightarrow \exists z {\rm Q}(\hat z,x)\\
{\overline {\rm Q}}(\hat x,y), {\rm P}(y) \rightarrow {\rm T}(\hat x).
\end{eqnarray*}
In the above set of rules, variable \(z\) in \({\rm Q}\) is existentially quantified
in the first rule. This implies that the first position of \({\rm Q}\) can bind to null
values (denoted by the ``hat'' symbol), 
for example in the second rule, where \(x\), also appears in the head.
Thus \(x\) is dangerous. However, since in the second rule it appears only
as a term of the predicate \({\rm Q}\) the ward (overlined). Also observe that \({\rm Q}\) shares \(y\),
a harmful variable, with \({\rm P}\).
\end{example}
Apart from Example~\ref{ex:simple_warded}, also 
the set of rules in Example~\ref{ex:company_and_kp} is warded, since although variable \(p\)
is dangerous in the second rule (as the first one can inject null values in the second position
of {\rm KeyPerson}), in the body it only appears in {\rm KeyPerson}, which is the ward 
(the ward is denoted by the overlining).}
Towards a formal definition of the notion of wardedness, let us first present some preliminary notions.
We consider a set of rules \(\Sigma\). For a predicate \(p\) used in \(\Sigma\), we define \emph{position} \(p[i]\) as the \(i\)-th term
of \(p\). Let \texttt{affected}\((\Sigma)\) be a set inductively as follows:
1. it contains all the positions \(\pi\) such that \(\pi\) is an existentially quantified variable for some rule of \(\Sigma\);
2. it contains all the positions \(\pi\), such that there is some rule \(\rho\) of \(\Sigma\)
where a variable \(v\) appears only in positions  \(\pi^\prime \in \mbox{\texttt{affected}}(\Sigma)\)
in the body of \(\rho\), and \(v\) appears in position \(\pi\) in the head of \(\rho\).
Vice versa, we define \texttt{nonaffected}\((\Sigma)\), the non-affected positions
of \(\Sigma\). When convenient, we denote variables in affected positions with
the ``hat'' symbol (\(\hat x\)).
In a given rule \(\rho\), the variables can be classified according to
their positions. We define a variable \(v\) as: \emph{harmless} in \(\rho\), if 
at least one occurrence in the body of \(\rho\) is in a non-affected position;
\emph{harmful} in \(\rho\), if in the body of \(\rho\), \(v\) always appears in affected positions;
\emph{dangerous}, if \(v\) is harmful in \(\rho\) and also appears in the head of \(\rho\).
In Example~\ref{ex:company_and_kp}, variable \(p\) is harmful and dangerous
(as the first rule can inject nulls into the second position
of {\rm KeyPerson}); \(x\) and \(y\) are harmless.

With these definitions in place, we define a set of rules as warded if the following conditions hold:
\emph{1. all the dangerous variables in a rule appear only within a single atom (the ward); 2. 
the ward shares with other atoms only harmless variables.}

\shortpaper{The set of rules in Example~\ref{ex:company_and_kp} is warded, since although variable \(p\)
is dangerous, in the body it only appears in {\rm KeyPerson}, which is the ward.}
\longpaper{
\noindent
In the following example we consider a more complex set of rules.

\begin{example}
\label{ex:complex_warded}
\begin{eqnarray*}
1: {\rm KeyPerson}(x,p) \rightarrow {\rm PSC}(x,p)\\
2: {\rm Company}(x) \rightarrow \exists p~{\rm PSC}(x,\hat p)\\
3: {\rm Control}(y,x), {\rm {\overline{PSC}}}(y,\hat p) \rightarrow {\rm PSC}(x,\hat p)\\
4: {\rm PSC}(x,\hat p), {\rm PSC}(y,\hat p), x>y \rightarrow {\rm StrongLink}(x,y) 
\end{eqnarray*}
\end{example}
Predicate {\rm PSC}\((x,p)\) asserts that \(p\) has
``significant control'' over company \(x\), in terms of decisions;
{\rm StrongLink}\((x,y)\) denotes somehow the presence of a strong connection between two companies.

\longpaper{Here the second rule introduces an existentially quantified variable \(p\) in {\rm PSC}.
In rule 3, \(p\) appears only in {\rm PSC} in affected position, so it is harmful.
Besides, it is dangerous since it appears in the head. However, {\rm PSC} is a ward,
since it contains all the dangerous variables and interacts with the rest of the body, namely
{\rm Company}, only through harmless variables, \(y\) in this case.
In the last rule, \(p\) is harmful, since it appears always in affected positions; however
it is not dangerous, because it does not appear in the head.}

The set of rules \shortpaper{in Example~\ref{ex:complex_warded}} respects wardedness and 
shows that it does not necessarily 
require the presence of a ward; for example, in the last rule, we accept a join on a harmful variable, 
\(p\) in this case. We will refer to this case as to \emph{harmful join}.}

\smallskip
\noindent
\cbstart
\new{\textbf{Other Datalog$^\pm$ Languages.} We now give a brief description of languages mentioned, but not used in this paper. In Linear Datalog$^\pm$, rule bodies are required to consist of only one atom. Guarded Data\-log$^\pm$
requires all universally quantified variables to occur within a single body atom. Sticky Datalog$^\pm$ and Weakly Sticky Datalog$^\pm$ pose syntactic  
restrictions based on a variable marking procedure. Weakly Frontier Guarded Datalog$^\pm$ tames the propagation of nulls by syntactic restrictions. In that sense, it is similar (but less strict) than Warded Datalog$^\pm$: It can be obtained from Warded Datalog$^\pm$ by dropping the requirement that the ward shares with other atoms only harmless variables.
} \cbend More details
can be found in \cite{GoPi15}.

\shorten{
\textbf{Modeling Features.}
We extend the core with a set of additional modeling features to support complex domains without affecting tractability.
\longpaper{We allow \emph{negative constraints} of the form 
$\forall \bar x (\varphi(\bar x) \rightarrow \bot),$
where $\varphi$ is a conjunction of atoms, and $\bot$ denotes the truth constant $\mathit{false}$
to model disjointness or non-membership; we support} 
\emph{equality-generating dependencies} (egds)~\cite{CaGP12} of the form
$\forall \bar x (\varphi(\bar x) \rightarrow x_i = x_j),$
where $\varphi$ is a conjunction of atoms, and $x_i, x_j$ are variable of $\bar x$,
which we assume not to interact with existential rules,
therefore preserving  the decidability of our 
reasoning task~\cite{ChVa85}.
}

\longpaper{The {\sc vadalog} language provides a simple mechanism to restrict the
possible bindings for body variables. We assume the presence of a unary
\emph{active constant domain} relation {\rm ACDom}~\cite{GoRS14}, 
defined as follows: for every database \(D\), 
{\rm ACDom}\((c) \in D\) iff \(c\) occurs in some atom \(R(\bar x) \in D\), with \(R \neq {\rm ACDom}\).
From the {\rm ACDom} relation, for each body rule \(\varphi(\bar x)\),
we define the n-ary \({\rm Dom}(*)\) relation as the cross product
\({\rm ACDom}(x_1) \times \ldots \times {\rm ACDom}(x_n)\), for all \(x_i \in \bar x\),
where \(n\) is the number of distinct variables in \(\varphi(\bar x)\).
The \({\rm Dom(*)}\) relation allows to restrict the 
possible bindings of the body variables to the ground values from the EDB.
The following example shows our modeling features in action.}
\longpaper{\begin{example}
\label{ex:dom}
\begin{eqnarray*}
1: {\rm \overline{Own}}(\hat x,y, \hat w) \rightarrow {\rm \underline{SoftLink}}(x,y)\\
2: {\rm SoftLink}(x,y) \rightarrow {\rm \underline{SoftLink}}(y,x)\\
3: {\rm Own}(\hat z,x,\hat{w1}), {\rm Own}(\hat z,y, \hat{w2}) \rightarrow {\rm \underline{SoftLink}(x,y)}\\
4: {\rm Incorp}(x,y) \rightarrow \exists z\exists w1\exists w2~{\rm Own}(\hat z,x,\hat{w1}), ~{\rm Own}(\hat z,y,\hat{w2})\\
5: {\rm Dom}(*), {\rm Incorp}(y,z), {\rm Own}(x1,y, w1), {\rm Own}( x2,z, w1) \rightarrow x1=x2\\
6: {\rm \overline{Own}}(\hat x,x, \hat w) \rightarrow \bot \\
\end{eqnarray*}
The goal of this reasoning task is determining the {\rm SoftLink} relationship between two companies.
Whenever a company \(x\) owns a company \(y\), the two are linked (and the relationship is symmetric).
If a third company \(z\) owns both \(x\) and \(y\), then the two are linked. An incorporation event
({\rm Incorp}) of two companies \(y\) and \(z\) means that there is a third company, \(x\), which
owns both. Rules 5 and 6 are constraints: if an incorporation takes place between \(y\) and \(z\),
there must be a unique single owner \(x\) of both. Rule 6 ensures that there is
no company that is self-owned.
\end{example}

\noindent
Observe that in Example~\ref{ex:dom}, we use {\rm Dom(*)} to restrict the application 
of rule 5, so that the egd constraint is never checked against non-ground values, like the ones
which could originate from rule 4. 
We also remark that the adoption of {\rm Dom(*)} does not affect either
decidability or tractability of the language.}

\section{Termination and recursion\\ control}
\label{sec:termination}

\shortRemove{Most reasoning tasks for Datalog with existential quantifiers are undecidable. In fact, the chase procedure for Datalog with existential quantification will in general not terminate due to the possibility of producing unboundedly many labelled nulls. However, the core of \textsc{Vadalog}, namely Warded Datalog\(^\pm\) is known to be decidable and capturing \textsf{PTIME}. Yet, this theoretical tractability does not immediately yield a practical algorithm. Specifically, the original proof for Warded Datalog\(^\pm\) used an alternating Turing machine, leaving no direct implementation strategy.}
\noindent
In this section, we present the core of the Vadalog system -- a practical algorithm with termination guarantees, yet allowing high performance and a limited memory footprint. This algorithm exploits the deep structural properties of Warded Datalog\(^\pm\), so our main focus in this section is to describe the key ideas behind the algorithm. In Section~\ref{sec:architecture}, describing the technical architecture,
we will show how the algorithm is actually exploited in a processing pipeline.

\cbstart
\new{Let us now consider a set of rules, based on a significant portion of the real-life
company control scenario (only advanced features have been omitted), 
already touched on in Example\shortRemove{s~\ref{ex:running} and}~\ref{ex:company_and_kp}, 
which we will use throughout the next sections as a running example. In particular, here we consider the
``significantly controlled companies'', that is, the companies for which there exist significant shareholders who hold
more than 20\% of the stocks.}
\begin{example}
\new{Assume we have the following database instance and set of rules, which are textually
described afterwards.}
\label{ex:running2}
\begin{eqnarray*}
\new{D = \{{\rm Company}({\rm HSBC}),{\rm Company}({\rm HSB}),{\rm Company}({\rm IBA}),}\\
\vspace{-4mm}
\new{
{\rm Controls}({\rm HSBC},{\rm HSB}),{\rm Controls}({\rm HSB},{\rm IBA})\}}\\[3mm]
\new{1: {\rm Company}(x) \rightarrow \exists p\exists s~{\rm Owns}(\hat p,\hat s,x)}\\
\new{2: {\rm Owns}(\hat p,\hat s,x) \rightarrow {\rm Stock}(x,\hat s)}\\
\new{3: {\rm Owns}(\hat p,\hat s,x) \rightarrow {\rm PSC}(x,\hat p)}\\
\new{4: {\rm PSC}(x,\hat p), {\rm Controls}(x,y) \rightarrow \exists s~{\rm Owns}(\hat p,\hat s,y)}\\
\new{5: {\rm PSC}(x,\hat p), {\rm PSC}(y,\hat p) \rightarrow {\rm \underline{StrongLink}}(x,y)}\\
\new{6: {\rm StrongLink}(x,y) \rightarrow \exists p\exists s~{\rm Owns}(\hat p,\hat s,x)}\\
\new{7: {\rm StrongLink}(x,y) \rightarrow \exists p\exists s~{\rm Owns}(\hat p,\hat s,y)}\\
\new{8: {\rm Stock}(x,\hat s) \rightarrow {\rm Company}(x)}.\\
\end{eqnarray*}
\new{
For a significantly controlled company $x$, there exists a person $p$ 
who owns significant shares $s$ of it (rule 1). The predicate {\rm Owns} denotes ``significant ownership''. 
When $p$ owns $s$\% of $x$, then $s$ is part of the share split of the company stock (rule 2). 
If $p$ owns a significant share of $x$, then $p$ is a ``person of significant control'' for $x$ (rule 3).
${\rm Controls}(x,y)$ holds when a company $x$ exerts some form of significant control over
company $y$; if ${\rm Controls}(x,y)$ holds, for every person $p$ having significant control 
over $x$, then there exist some share $s$ ($s>20$\%) , 
such that $p$ owns $s$\% of $y$ (rule 4). 
If two companies share a person with significant control, they have a strong link (rule 5). 
If there is a strong link between $x$ and $y$, 
there is some person $p$ who owns significant share $s$ of $x$ (rule 6); 
the same applies to $y$ (rule 7). Finally, the existence of a stock for $x$, implies that $x$ 
is a company (rule 8). We are interested in knowing all the strong links between companies (underlined).
}
\end{example}
\cbend

\subsection{Warded Forest}
\label{sec:warded_forest}
\noindent
The key for guaranteeing termination of a chase-based procedure is defining points at which the chase procedure may be terminated prematurely, while at the same time upholding correctness of the reasoning task, i.e., guaranteeing that all tuples that are part of the output have already been produced. We shall call a principle by which such ``cut-off'' of the chase tree can be achieved a \textit{termination strategy}. Generally speaking, the earlier this cut-off can be achieved, the better our termination strategy. In particular, for it to be effective, it has to always cut-off after a finite part of the, in general, unboundedly large chase graph. Yet at the same time, the overhead of the strategy itself has to be taken into consideration.

More formally speaking, we define the \emph{chase graph} for a database \(D\) and a set of rules \(\Sigma\), as the directed
graph $\textit{chase-graph}(D,\Sigma)$ having as nodes the facts obtained from \(\mbox{chase}(D,\Sigma)\) 
and having an edge from a node {\bf a} to {\bf b} if {\bf b} is obtained from {\bf a} and possibly from
other facts by the application of one chase step, i.e., of one rule of \(\Sigma\).
\medskip

As the main vehicle towards understanding the chase graph for Warded Datalog$^\pm$, we introduce our first major tool, namely
the \textit{warded forest} of a chase graph: the subgraph that consists of all nodes of the chase graphs, all edges of the chase graph  that correspond to the application of linear rules, and one edge for each non-linear rule -- namely the one from the fact bound to the ward.
Thus, by this definition, the connected components in a warded forest are explicitly determined (i.e., separated from each other) by joins involving constants
or, in other words, in each connected component we only have edges representing the application of linear rules or rules with
warded dangerous variables. As an effect, every single fact will then inherit all its labelled nulls
from exactly one fact, either its direct parent in the application of a linear rule or the ward
in the application of a warded rule.

\cbstart
\new{Figure~\ref{fig:warded_forest_example} shows a portion of the (infinite) warded forest
of the chase graph for our running example (Example~\ref{ex:running2}). In particular, solid edges
derive from the application of linear rules, and dash-dotted edges derive
from non-linear rules where one fact involved in the join
binds to a ward. The gray rectangles denote the subtrees in the warded forest:
they are generated by non-linear rules where none of the facts involved in the
join bind to a ward, i.e., there is no propagation of dangerous variables to the head.}
\cbend

\old{The example in Figure~\ref{fig:forest_example} shows a set of warded rules \(\Sigma\), a portion of the respective (infinite)
chase graph considering all the edges in the figure, and a portion of the (infinite) warded forest. In particular,
solid edges derive from the application of linear rules whereas dashed-dotted ones derive from the application
of non-linear rules where one fact involved in the join binds to a ward. Finally, dotted edges denote the application of a non-linear
rule, where none of the facts involved in the join bind to a ward, i.e., there is no propagation of dangerous variables to the head.
The gray rectangles denote the subtrees in the warded forest.}

\old{\begin{figure*} [t!]
\centering
 \begin{minipage}{\linewidth}
\begin{minipage}{0.45\linewidth}
\vspace{-6mm}
\begin{eqnarray*}
D = \{{\rm P}(a), {\rm Q}(a,c)\}\\[3mm]
1: {\rm P}(x) \rightarrow \exists z~{\rm \underline{Q}}(x,\hat z)\\
2: {\rm Q}(x,\hat y) \rightarrow {\rm S}(\hat y, x)\\
3: {\rm S}(\hat x, y),{\rm P}(y) \rightarrow {\rm T}(y, \hat x)\\
4: {\rm T}(x, \hat y),{\rm Q}(z, \hat y) \rightarrow {\rm \underline{H}}(x,z)\\
5: {\rm T}(x,\hat y) \rightarrow {\rm \underline{Q}}(x,x)\\
6: {\rm Q}(x,\hat y) \rightarrow {\rm \underline{F}}(\hat y)\\
7: {\rm H}(x,x) \rightarrow \exists z~{\rm \underline{Q}}(x,\hat z)\\
8: {\rm P}(x) \rightarrow \exists z~{\rm T}(x,\hat z).
\end{eqnarray*}
\end{minipage}
\begin{minipage}{\linewidth}
\hspace{-4mm}
    \includegraphics[scale=0.30]{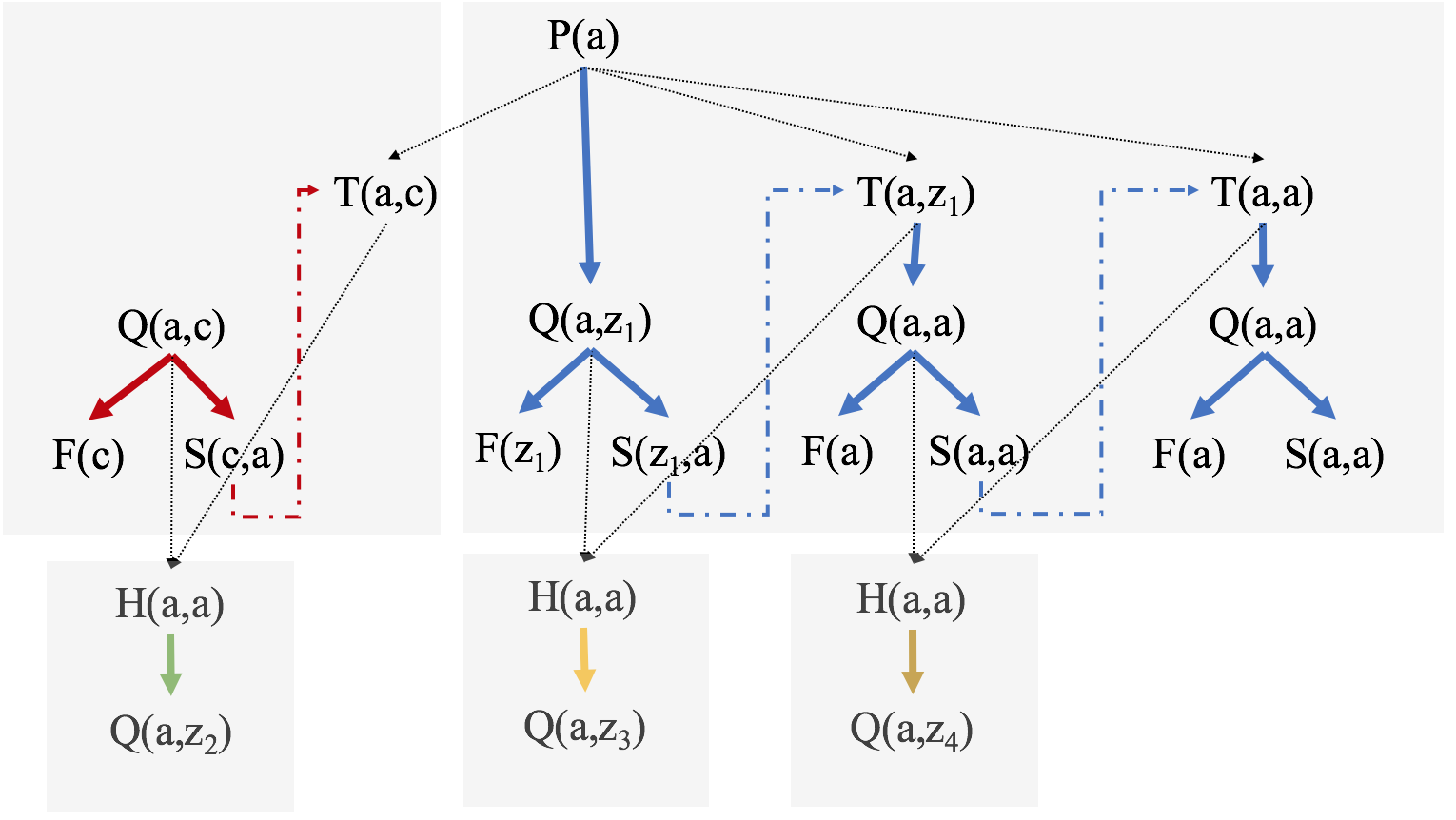}
\end{minipage}

\end{minipage}
\caption{On the left, a set of warded rules \(\Sigma\) for a database $D$. On the right, the respective structures: the chase graph for ${\rm chase}(\Sigma, D)$ (considering all the edges); the warded forest (considering the solid and dashed-dotted edges); the linear forest (considering the solid edges).}
\label{fig:forest_example}
\end{figure*}}

We say that two facts are \emph{isomorphic} if they refer to the same predicate name, have the same constants in the same positions and 
there exists a bijection of labelled nulls into labelled nulls.
Then, we define a \emph{subtree}({\bf a}) as the subtree of the warded forest that is rooted in {\bf a}. We say that two 
subtrees are isomorphic if the usual notion of graph isomorphism holds with the given definition of
isomorphism between facts.

\begin{theorem}
\label{th:isomorphism}
Let {\bf a} and {\bf b} be two facts in a warded forest. If they are isomorphic, then 
\emph{subtree}({\bf a}) is isomorphic to  \emph{subtree}({\bf b}).
\longpaper{\begin{proof}
Let us assume {\bf a} and {\bf b} are isomorphic. We show inductively that \emph{subtree}({\bf a}) is also 
isomorphic to \emph{subtree}({\bf b}). 
Let \({\bf a}^{\prime\prime}\) be a fact in \emph{subtree}({\bf a}), obtained at chase
step \(i\) and let us also pose (inductive hypothesis) that all the facts
generated for \emph{subtree}({\bf a}) in chase steps \(<i\) are isomorphic to corresponding facts in \emph{subtree}({\bf b}),
which has already been fully generated by the chase.
Towards a contradiction, let us assume that \({\bf a}^{\prime\prime}\) is not isomorphic to any fact in \emph{subtree}({\bf b})
(and therefore \emph{subtree}({\bf a}) is not isomorphic to \emph{subtree}({\bf b})).
Let \({\bf a}^\prime\) in \emph{subtree}({\bf a}) be the fact from which \({\bf a}^{\prime\prime}\) is generated.
Since by inductive hypothesis for every chase step \(<i\) the generated facts in the two subtrees are isomorphic, there
is an fact \({\bf b}^\prime\) in \emph{subtree}({\bf b}) isomorphic to \({\bf a}^\prime\).
As they are isomorphic, any linear rule that applies to \({\bf a}^\prime\) also applies to \({\bf b}^\prime\) and vice versa.
This means that the only way to produce \({\bf a}^{\prime\prime}\) such that it is not isomorphic to any fact in \emph{subtree}({\bf b}),
is that the chase generates it by applying a non-linear rule \(\rho\) that joins \({\bf a}^\prime\) with another fact
\({\bf k}\) that shares a labelled null \(\nu\) with  \({\bf a}^\prime\), while \(\nu\) does not appear in (the same position of)
\({\bf b}^\prime\).
Since \(\nu\) appears as an argument of both \({\bf a}^\prime\) and \({\bf k}\), then two cases are possible: 
1. the chase has ``transferred'' \(\nu\) from \({\bf a}^\prime\) to \({\bf k}\) (or vice versa) by applying only linear rules.
This is impossible, as \({\bf b}^\prime\) is isomorphic to \({\bf a}^\prime\) and so the same sequence of linear rules 
would be applicable also from \({\bf b}^\prime\) to {\bf k} to (or vice versa) and therefore \(\nu\) would appear
in \({\bf b}^\prime\) as well; 
2. \(\nu\) in \({\bf a}^\prime\) derives from another fact \({\bf a}^0\) (of \emph{subtree}({\bf a}) by definition of warded forest), 
which the chase joined at some step \(<i\)
with another fact \({\bf m}\) with a non-linear warded rule \(\rho^\prime\). Since \(\rho^\prime\)
propagates a labelled null from the body (in particular from \({\bf a}^0\)) to the head (here to \({\bf a}^\prime\)), 
by wardedness, \({\bf m}\) interacts with \({\bf a}^0\) only via harmless variables, i.e., 
constants; since, by inductive hypothesis, there exists a fact \({\bf b}^0\) of \emph{subtree}({\bf b}), isomorphic to
\({\bf a}^0\), rule \(\rho^\prime\) would also join \({\bf m}\) and \({\bf b}^0\) transferring  \(\nu\) to \({\bf b}^\prime\)
and thus violating the fact that \(\nu\) does not appear in (the same position of)
\({\bf b}^\prime\).
The contradiction of conditions 1 and 2 shows that \emph{subtree}({\bf a}) is isomorphic to  \emph{subtree}({\bf b}).
\end{proof}}
\end{theorem}

\noindent In the warded forest in \old{Figure~\ref{fig:forest_example}}\new{Figure~\ref{fig:warded_forest_example},} \cbstart\new{we see examples of isomorphic facts, which give rise
to isomorphic subtrees (each shown only once):} \old{those rooted in ${\rm H}(a,a)$,
the ones rooted in ${\rm Q}(a,a)$, and so on.}\new{those rooted in ${\rm Company}({\rm HSBC})$, in ${\rm Owns}(p,s,{\rm HSBC})$, and so on.} \cbend 
Theorem~\ref{th:isomorphism} gives a fundamental result in terms of description of the ``topology'' of the warded
chase forest, since we see that each subtree is uniquely identified, up to isomorphism of the
\shortRemove{labelled}nulls, by its root. It is certainly an important hint for the exploration of the chase graph, as it points
out a form of structural periodicity that we will exploit to efficiently guarantee termination. 
\shortRemove{Note that apart from its role of pointing towards this structural periodicity, the warded forest underlying Theorem 1 will also play a very interesting role in its own right, concerning improving performance and limiting the space footprint of the chase.}

\cbstart
\begin{figure}
\centering
    \includegraphics[scale=0.3]{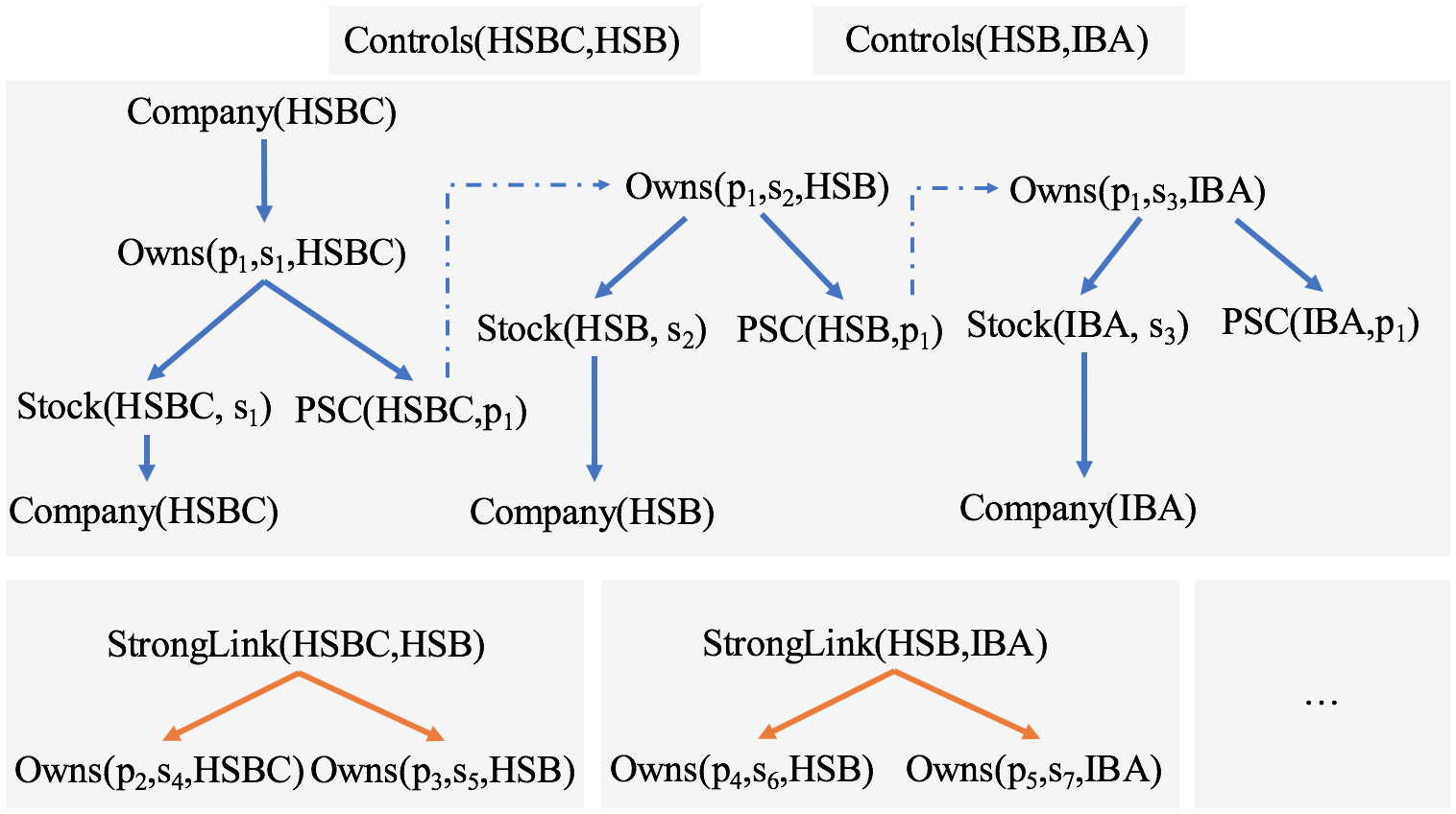}
\vspace{1ex}\caption{\newspecial{A portion of the warded forest for the chase graph of rules in Example~\ref{ex:running2}.
Solid edges denote linear rules and the dash-dotted edges denote
joins in warded rules.}}
\label{fig:warded_forest_example}
\end{figure}
\cbend

\subsection{Harmless Warded Datalog\(^\pm\)}
\label{sec:harmless}

\noindent
We now make a step forward towards the definition of a concrete termination strategy in the 
application of the chase. Theorem~\ref{th:isomorphism} gives a good direction, since it ensures that
for isomorphic facts, subtrees rooted at these facts will be isomorphic as well. Therefore, ideally, when the chase produces a fact
that is isomorphic to one that has been generated at a previous step, we could avoid exploring its successors, since the
derived subtree would certainly be isomorphic to an already generated one and therefore irrelevant for our
reasoning task. Unfortunately, this property of subtrees in warded forests does not extend to generic chase graphs
(and graph isomorphism), as shown by the following example.

\cbstart
\begin{example}
\label{ex:harmful_join}
\new{From database $D$ of Example~\ref{ex:running2}, 
we can generate a fact ${\rm Owns}(p_1,s_1,{\rm HSBC})$ with rule 1, 
and $Owns(p_1,s_2,{\rm HSB})$ by subsequently applying rule 3 (which produces
${\rm PSC}({\rm HSBC},p_1)$) and rule 4.
By applying rule 3 on $Owns(p_1,s_2,{\rm HSB})$, we also obtain ${\rm PSC}({\rm HSB},p_1)$. 
From ${\rm Company}({\rm HSB})$, 
via rule~1 and rule~3, we obtain ${\rm PSC}({\rm HSB},p_2)$. 
Then, we see that although ${\rm PSC}({\rm HSB},p_1)$ and ${\rm PSC}({\rm HSB},p_2)$ 
are isomorphic, the respective subtrees are not: 
${\rm PSC}({\rm HSB},p_1)$ has ${\rm StrongLink}({\rm HSBC},{\rm HSB})$ 
as a child node, which derives from the join with 
${\rm PSC}({\rm HSBC},p_1)$ in rule 5; conversely, 
${\rm PSC}({\rm HSB},p_2)$ does not join with any ${\rm PSC}$ facts that regard ${\rm HSBC}$.
}\end{example}
\cbend
\old{\begin{example}
\label{ex:harmful_join}
$1: {\rm R}(x) \rightarrow \exists z~{\rm T}(\hat z)$
$2: {\rm R}(x) \rightarrow \exists z~{\rm P}(\hat z,x)$\\
$3: {\rm P}(\hat z,x) \rightarrow {\rm T}(\hat z)$
$4: {\rm P}(\hat z,x),{\rm T}(\hat z) \rightarrow {\rm \underline{Q}}(x).$

\smallskip \noindent 
Considering the database \(D=\{{\rm R}(a)\}\), we can generate a fact \({\rm T}(\bot_1)\) via rule 1
and a fact \({\rm T}(\bot_2)\) by subsequently applying rule 2 (which generates an intermediate fact \(P(\bot_2, a)\)) and rule 3. Then, although \({\rm T}(\bot_1)\) and \({\rm T}(\bot_2)\) are isomorphic, the 
respective subtrees are not isomorphic: \({\rm T}(\bot_2)\) has \({\rm Q}(a)\) as a child node, which derives from the join with
with \(P(\bot_2, a)\) in rule 4, whereas \({\rm T}(\bot_1)\) does not join with any facts for predicate \({\rm P}\).
\end{example}}

\noindent
Example~\ref{ex:harmful_join} shows that pruning a chase graph on the basis of isomorphism on
labelled nulls is not correct. However, if we consider a 
syntactical restriction of Warded Datalog\(^\pm\), namely \emph{Harmless Warded Datalog}\(^\pm\),
where joins on harmful variables are forbidden, then Theorem~\ref{th:isomorphism} 
extends to generic chase graphs.

\begin{theorem}
\label{th:isomorphism_harmless}
Let {\bf a} and {\bf b} be two facts in the chase graph of a set of
harmless warded rules. If {\bf a} and {\bf b} are isomorphic,
then \emph{subgraph}({\bf a}) is isomorphic to \emph{subgraph}({\bf b}).
\longpaper{\begin{proof}
Intuitively, the only way to have non-isomorphic subtrees is that some
successor \({\bf a}^{\prime\prime}\) of {\bf a} joins with a fact {\bf k} in the graph,
which does not join with any fact \({\bf b}^{\prime\prime}\), successor of {\bf b} and isomorphic
to \({\bf a}^{\prime\prime}\). If the join is on harmless variables this is not possible, as the same constant
would appear in both \({\bf a}^{\prime\prime}\) and \({\bf b}^{\prime\prime}\) since they are isomorphic.
If the join is on harmful variables, it could be the case that 
a labelled null \(\nu\) appears in \({\bf a}^{\prime\prime}\) but not in
\({\bf b}^{\prime\prime}\) and hence {\bf k} joins only with \({\bf a}^{\prime\prime}\). Yet the presence
of joins on harmful variables is forbidden by definition of Harmless Warded Datalog.
\end{proof}}
\end{theorem}

\noindent
\longpaper{Theorem~\ref{th:isomorphism_harmless}} immediately suggests a (typically not very efficient) terminating algorithm for Harmless Warded Datalog$^\pm$, namely by modifying the chase in two aspects: 

\vsRemove{\begin{itemize}
\item \textit{detection}: memorizing all atoms generated during the chase up to isomorphism;
\item \textit{enforcement}: cutting off the chase when an atom isomorphic to a memorized one would be generated.
\end{itemize}}

Correctness of the algorithm follows from Theorem~\ref{th:isomorphism_harmless} and the fact that our reasoning task is not affected by the multiplicity of isomorphic copies of facts. Yet, this algorithm is of course impractical, as memorizing the chase up to isomorphism and performing the full isomorphism check is prohibitive in space and time \cbstart
\new{as we will show experimentally in Section~\ref{sec:homomorphism}}.
\cbend

A limitation of Theorem~\ref{th:isomorphism_harmless}, but only an apparent one, is that is restricted to Harmless Warded Datalog$^\pm$. Yet, it fully extends to Warded Datalog\(^\pm\) as
a set of warded rules can be always rewritten as an equivalent (i.e., giving the same result for any instance of our reasoning task up to null homomorphism) set of harmless warded rules.
For this purpose we devise a \emph{Harmful Joins Elimination Algorithm}, which we present next. It is inspired, in part, by work on schema mappings  \cite{tods/FaginKPT05,pods/KolaitisPSS14,mst/KolaitisPSS18,mst/PichlerSS13,dagstuhl/Sallinger13}.
\cbstart
\shortRemove{

\medskip\noindent
{\em Harmful Joins Elimination Algorithm:} 
Let us consider a set \(S\) of warded rules, where some of them contain harmful joins (\emph{harmful rules}).
Without loss of generality, such rules can always be written (by breaking
more complex rules into multiple steps) as having exactly one harmful join, in the following form:
 \[\alpha: \forall \bar x \forall \bar y \forall h ({\rm A}(\bar x_1, \bar y_1, \hat h), {\rm B}(\bar x_2, \bar y_2, \hat h) \rightarrow  \exists\bar z~{\rm C}(\bar x, \bar z)) \]
 where A, B and C are predicate names,
\(\bar x_1, \bar x_2 \subseteq \bar x\), \(\bar y_1, y_2 \subseteq \bar y\) are (possibly empty) disjoint sets of harmless
variables or constants and \(h\) is a harmful variable. An example is rule 4 in
Example~\ref{ex:harmful_join}.
The goal of the following algorithm is to remove harmful rules.\\

\noindent Until there are harmful rules \(\alpha\) in \(S\) (\emph{cause elimination}):
\begin{compactenum}
\item \emph{Grounding:} Add to \(S\) the following rules:\\ 
\( {\rm Dom}(h), {\rm A}(\bar x_1, \bar y_1, \hat h) \rightarrow {\rm A}^\prime(\bar x_1, \bar y_1, \hat h)\)

\({\rm A^\prime}(\bar x_1, \bar y_1, \hat h), {\rm B}(\bar x_2, \bar y_2, \hat h) \rightarrow  \exists \bar z~{\rm C}(\bar x, \bar z) \)

\item \emph{Direct:} For each rule \(\beta: \forall \bar x \forall \bar y ~\varphi(\bar x, \bar y) \rightarrow  \exists \bar z \exists h~A(\bar x, \bar z, \hat h) \), add to \(S\):
\(\varphi(\bar x_1, \bar y_1), {\rm B}(\bar x_2, \bar y_2, f_\beta(\bar x_1, \bar y_1)\new{)}  \rightarrow  \exists \bar z~{\rm C}(\bar x, \bar z) \)

\item \emph{Indirect:} For each rule \(\beta: \varphi(\bar x, \bar y, \hat h) \rightarrow  \exists \bar z ~A(\bar x, \bar z, \hat h) \), add to \(S\) the rule: 
\(\varphi(\bar x_1, \bar y_1, \hat h), {\rm B}(\bar x_2, \bar y_2, \hat h)  \rightarrow  \exists \bar z~{\rm C}(\bar x, \bar z) \)

\item remove \(\alpha\) from \(S\).\\
\end{compactenum}

\noindent For each rule \(\rho\) added to \(S\) at the previous step \(S\), depending on its
form, do as follows (\emph{Skolem simplification}):

\begin{compactenum}
\item \emph{Virtual Joins:} Remove from \(S\) rules \(\rho\) of the form: \\
a.~\({\rm A}(\bar x_1, \bar y_1, h), {\rm B}(\bar x_2, \bar y_2, f_\beta(h)),\ldots \rightarrow  \exists\bar z~{\rm C}(\bar x, \bar z) \)\\
b.~\({\rm A}(\bar x_1, \bar y_1, h=f_{\beta_1}(\cdot)), {\rm B}(\bar x_2, \bar y_2, h=f_{\beta_2}(\cdot)), \ldots \rightarrow  \exists\bar z~{\rm C}(\bar x, \bar z) \)\\
c.~\({\rm A}(\bar x_1, \bar y_1, h=f_{\beta}(\cdot)), {\rm B}(\bar x_2, \bar y_2, h=f_{\beta}(f_\beta (...\cdot))), \ldots \rightarrow  \exists\bar z~{\rm C}(\bar x, \bar z) \)
\item \emph{Linearization:} Unify into \(S\) rules \(\rho\) of the form:\\
\({\rm A}(\bar x_1, \bar y_1, h=f_\beta(\cdot)), {\rm A}(\bar x_2, \bar y_2, h=f_\beta(\cdot)), \ldots \rightarrow  \exists\bar z~{\rm C}(\bar x, \bar z) \), into\\ \({\rm A}(\bar x, \bar y, h),\ldots \rightarrow  \exists\bar z~{\rm C}(\bar x, \bar z). \)\\
\end{compactenum}

\noindent
The cause elimination part takes care of removing harmful rules from \(S\). In the grounding
we produce a ground harmless copy of \(\alpha\), which operates only on EDB facts by
using the {\rm Dom} feature (recall that {\rm Dom} ensures that variables in {\rm Dom} bind only to constants in the domain, and not to nulls). Then for each harmful variable \(h\) involved in a harmful
join in an atom \(A\), we individuate all the ``causes'', i.e., the rules \(\beta\) whose head
unifies with \(A\) and propagate labelled nulls that then bind to \(h\). This
propagation can be direct (\(\beta\) has an existential quantification on \(h\)) or
indirect (\(\beta\) propagates a labelled null originally received from other
rules). We rewrite \(\alpha\) by replacing \(A\) with the body of \(\beta\) according
to usual composition semantics~\cite{FKPT11}. In particular, we account for
the existential quantification in the rewriting by introducing Skolem functions \(f_\beta\).
We perform this process iteratively as long as \(S\) contains harmful joins.
It is intuitive to see that thanks to the definition of Warded Datalog\(^\pm\),
this process is bound to terminate and that the number of generated rules
is in worst case exponential in the number of the original harmful ones.

The Skolem simplification part takes care of managing the rules that contain 
Skolem functions. In three cases, these rules can be dropped as the conditions
on the functions cannot be satisfied (and we say that the joins are virtual). Specifically, this happens
when: (1a) a harmless variable is equated to a Skolem function, impossible since a labelled null differs from any constant; (1b) 
two distinct Skolem functions are equated, impossible since ranges are disjoint by definition; (1c) a 
Skolem function is equated to its recursive application, impossible since Skolem functions are
injective.
Finally, in the linearization case, a rule involving Skolem functions can be simplified in such a way that two atoms
with the same Skolem function are unified, and the function is replaced by a variable.

The following example shows \old{the result of the application of the harmful joins elimination
algorithm to the set of rules in Example~\ref{ex:harmful_join}.}\new{the elimination
of harmful join in rule 5 in Example~\ref{ex:running2}.
Note that we report only the relevant portion of the resulting set of rules.}}

\old{\begin{example}
\label{ex:harmful_join_simplified}
${\rm 1: R}(x) \rightarrow \exists z~{\rm T}(\hat z)~~$
${\rm 2: R}(x) \rightarrow \exists z~{\rm P}(\hat z,x)~~$\\
\hspace*{2em}${\rm 3: P}(\hat z,x) \rightarrow {\rm T}(\hat z)~~$
${\rm 4: Dom}(z), {\rm P}(z,x),{\rm T}(z) \rightarrow {\rm \underline{Q}}(x)$~~\\
\hspace*{2em}${\rm 5: R}(x) \rightarrow {\rm \underline{Q}}(x).$

\smallskip \noindent
Rule 4 is the grounded version of rule 4 in Example~\ref{ex:harmful_join}.
By repeatedly composing the original rule 4 with rules 1 and 2 (which inject
labelled nulls into \(y\) of rule 4), we obtain rule 5 by linearization.
\end{example}}

\begin{example}
\label{ex:harmful_join_simplified}
\new{
${\rm Dom}(p), {\rm PSC}(x,p) \rightarrow {\rm PSC}^\prime(x,p)~~$\\
${\rm PSC}^\prime(x,p), {\rm PSC}^\prime(y,p) \rightarrow {\rm StrongLink}(x,y)~~$\\
${\rm Company}(x), {\rm Controls}(x,y) \rightarrow {\rm StrongLink}(x,y), {\rm StrongLink}(y,x)$~~\\
${\rm Company}(x), {\rm Controls}(x,z), {\rm StrongLink}(z,y) \rightarrow$ \\
\indent\indent\indent\indent\indent\indent\indent\indent ${\rm StrongLink}(x,y), {\rm StrongLink}(y,x).$
}
\medskip \noindent
\end{example}

\noindent
\new{Essentially, the first two rules encode the grounded version of the original rule (i.e., one only considering constants).
They are obtained by creating a ground copy of the predicate holding the harmful variable.
The last two rules encode transitive closure of the StrongLink relation. They are obtained by means of a rewriting-based technique, which replaces atoms that contain harmful variables. Thanks to the 
definition of Warded Datalog\(^\pm\), this procedure always terminates.
}
\cbend

\subsection{Lifted Linear Forest}
\label{sec:linear_forest}

\noindent
In Section~\ref{sec:warded_forest} we have seen how Theorem~\ref{th:isomorphism} guarantees
isomorphism of subtrees in warded forests based on the isomorphism of their roots.
In Section~\ref{sec:harmless} we showed how this result can be extended to chase
graphs (Theorem~\ref{th:isomorphism_harmless}) if the rules do not contain harmful
joins (Harmless Warded Datalog\(^\pm\))\vsRemove{and we presented an algorithm (Harmful Join Elimination)
to construct harmless warded rules from warded rules}. 

Ideally this would be enough for the implementation of a termination strategy, since the chase
could skip the exploration of subgraphs isomorphic to other nodes that have already
been explored (\emph{vertical pruning}). This would however have practicability limitations
in that the algorithm would rely on the specific ground values in the facts to deduce the isomorphism 
of the derived subgraphs with the result of very low chances to detect isomorphic subtrees.

In this section we introduce the notion of \emph{lifted linear forest}, a structure that allows
to apply the results on isomorphisms independently of the specific ground values by grouping
subtrees with the same ``topology'' into equivalence classes.

We say that two facts are \emph{pattern-isomorphic} if they have the same predicate name,
there exists a bijection between the constant values and there exists a bijection between the labelled nulls.
For example, \({\rm P}(1,2,x,y)\) is pattern-isomorphic to \({\rm P}(3,4,z,y)\), but not to \({\rm P}(5,5,z,y)\).
By extension, we define two graphs as pattern-isomorphic if the nodes are pairwise isomorphic and they
have coinciding edges. By means of pattern-isomorphism we can group the facts
into equivalence classes and we denote with \(\pi({\bf a})\) the representative of the equivalence class to which
{\bf a} belongs. In other terms, \(\pi\) is a function such that it maps pattern-isomorphic facts into
the same element, which we call \emph{pattern}; for example, it could map \({\rm P}(1,2,x,y)\) and 
\({\rm P}(3,4,z,y)\) into \({\rm P}(c_1,c_2,\nu_1,\nu_2)\) and \({\rm P}(5,5,z,y)\) into \({\rm P}(c_1,c_1,\nu_1,\nu_2)\), 
though any other representation for patterns would be acceptable. A notion of \emph{graph pattern-isomorphism} can be thus
easily derived by extension.

An immediate use would be grouping subgraphs of the chase graphs into equivalence classes,
so that Theorem~\ref{th:isomorphism_harmless} can be applied more broadly, deriving 
subgraph pattern-isomorphism on the basis of the pattern-isomorphism of the respective root nodes.
This would open to the implementation of a form of \emph{horizontal pruning}, i.e., 
vertical pruning modulo pattern-isomorphism. Unfortunately this is not possible: given two
pattern-isomorphic nodes {\bf a} and {\bf b}, the two derived subgraphs in the chase graph can be different indeed,
as one can have a fact \({\bf a^{\prime\prime}} \in {\rm \emph{subgraph}}({\bf a})\), pattern-isomorphic to \({\bf b^{\prime\prime}} \in {\rm \emph{subgraph}}({\bf b})\)
such that there is a third fact {\bf k} sharing a constant \(c\) with \({\bf a^{\prime\prime}}\) but not with \({\bf b^{\prime\prime}}\).
A non-linear rule could then join  \({\bf a^{\prime\prime}}\) with {\bf k}, producing \({\bf a^{\prime}}\). Since {\bf k} does not join
with \({\bf b^{\prime\prime}}\), in \({\rm \emph{subgraph}}({\bf b})\) there would be no facts pattern-isomorphic to  \({\bf a^{\prime}}\).
This means that in harmless warded forests, subgraph isomorphism is independent of the specific values of the labelled nulls but is dependent on the specific values of the constants. 

Towards overcoming this limitation, we introduce the notion of \emph{linear forest}.
A linear forest for a chase graph is the subgraph that contains all the facts from the graph
and only the edges that correspond to the application of linear rules (one atom in the body).

\old{We have an example of linear forest in Figure~\ref{fig:forest_example}, where it is
denoted by the subgraph where only the solid edges are considered.} \cbstart
\new{We have an example of linear forest in Figure~\ref{fig:warded_forest_example},
where it is denoted by the subgraph where only the solid edges are considered.} \cbend
In a linear forest, pattern-isomorphism on roots extends to pattern-isomorphism of subtrees,
as shown by the following property.

\begin{theorem}
\label{th:linear_forest}
Let {\bf a} and {\bf b} be two facts in a linear forest. If they are pattern-isomorphic, then
\emph{subtree}\(({\bf a})\) is pattern-isomorphic to \emph{subtree}\(({\bf b})\).

\longpaper{\begin{proof}
 We proceed inductively. Let us assume without loss of generality that the rules that generated
the forest do not have constants.
Let us assume that two facts 
\({\bf a}^{\prime} \in {\rm \emph{subgraph}}({\bf a})\) and  \({\bf b}^{\prime} \in {\rm \emph{subgraph}}({\bf b})\) are pattern isomorphic (inductive hypothesis).
Then, since within a subtree only linear rules are applied, we have
that the body of any linear rule unifies with  \({\bf a}^\prime\) iff it unifies with \({\bf b}^\prime\)
and therefore the conclusion \({\bf a^{\prime\prime}}\) of \({\bf a^\prime}\)
is pattern-isomorphic to the conclusion \({\bf b^{\prime\prime}}\) of \({\bf b^\prime}\)
and by induction the two subtrees are pattern-isomorphic.
\end{proof}}
\end{theorem}

\begin{figure}
\centering
    \includegraphics[scale=0.3]{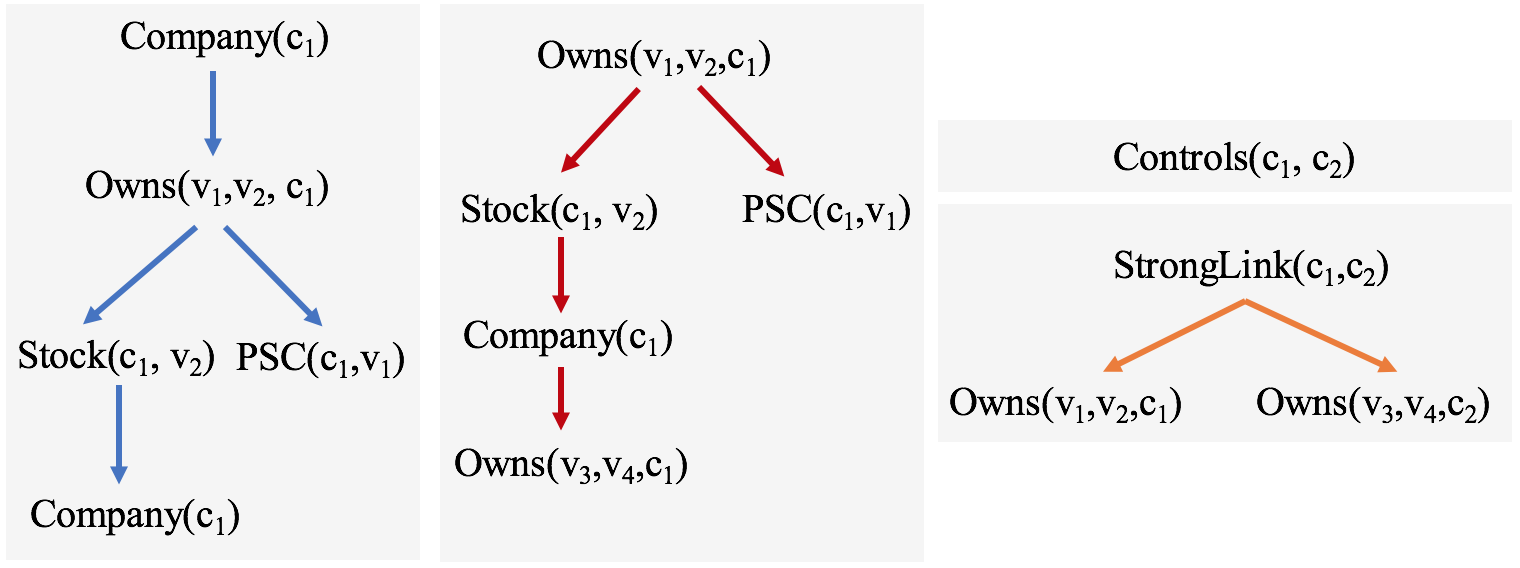}
\caption{\newspecial{The lifted linear forest for the chase graph of Example~\ref{ex:running2}.}}
\label{fig:lifted_linear_forest_example}
\end{figure}

\noindent Inspired by the application of Theorem~\ref{th:linear_forest} 
to the notion of linear forest, we define the \emph{lifted linear forest} for a chase graph 
as the set of equivalence classes modulo subtree pattern-isomorphism of the linear forest.
For each node {\bf n} of a linear forest, the corresponding node \(\pi({\bf n})\) belongs
to the lifted linear forest; the subtrees of the linear forest whose roots are pattern-isomorphic
are thus collapsed into the same connected component of the lifted linear forest.
The term ``lifted'' suggests that such data structure seizes the structural symmetries in the linear forests and factors them out.
\cbstart
\new{An example of lifted linear forest is presented in Figure~\ref{fig:lifted_linear_forest_example},
where we show how the chase graph for Example~\ref{ex:running2} is simplified.}
\cbend

In linear forests, a termination strategy could thus feature both vertical and horizontal pruning.
This however comes at a price: as trees in linear forests tend to be less deep than in warded forests because of the exclusion
of the edges for joins in warded rules, node isomorphisms are less beneficial, as they allow
to prevent fewer isomorphism checks. The next section shows how we combine
warded and lifted linear structures to achieve the best trade off.

\cbstart
\cbend

\subsection{The Algorithm}
\label{sec:algorithm}

\noindent
With all ingredients ready, we now present the main algorithm of the Vadalog system.
Algorithm~\ref{alg:termination} is a termination strategy in the sense that it
guides \({\rm chase}(D,\Sigma)\) and returns whether each chase step must be
activated or not. Thus it controls the generation of new facts and, hence, 
the termination of the (in general infinite) chase.

The core principle of it is incrementally building a warded forest as in Theorem~\ref{th:isomorphism_harmless}
and a lifted linear forest, and, at the same time, exploiting
the already available portions of these forests in multiple ways. Specifically, we will
use the warded forest for an optimized form of local detection of isomorphic facts and the lifted
linear for a highly reusable and compact summary of the detected patterns for their enforcement.
This termination strategy can be part of a na\"{i}ve chase algorithm, \shortRemove{as e.g. illustrated in Algorithm~\ref{alg:chase},}but more practically -- for achieving good performance -- in the Vadalog system is applied in the streaming-based pipeline described in Section~\ref{sec:architecture}.

\medskip \noindent
The algorithm uses a number of data structures based on one of the three main concepts introduced so far. In particular, the properties associated to facts (namely root and provenance) are mostly based on the concept of \textit{linear forest}. 
The ground structure, which is the target of isomorphisms checks, is based on the \textit{warded forest}. \cbstart \new{
We saw a full example of this in Figure~\ref{fig:warded_forest_example}}. \cbend Finally, a summary structure that determines when skipping isomorphisms checks is possible, is based on the \textit{lifted linear forest}. \cbstart \new{We saw an example of this in Figure~\ref{fig:lifted_linear_forest_example}}. \cbend
We describe each of them in detail now.

The \textbf{fact structure}
is a structured representation of a fact {\bf a}, with four fields. 
1. \emph{generating\_rule:} The kind of rule of \(\Sigma\) (linear / warded / non-linear) which generated 
{\bf a}. Here a ``warded'' rule is a rule where there is actually a join over a harmless variable in a ward
and a dangerous variable is propagated to the head.
2. \emph{l\_root:} The root of the tree (i.e., of the connected component) containing {\bf a} in the linear forest.
3. \emph{w\_root:} The root of the tree (i.e., of the connected component) containing {\bf a} in the warded forest.
4. \emph{provenance:} The provenance of {\bf a} in the linear forest,
defined as the list $[\rho_1, \ldots, \rho_n]$ 
of the rules that have been orderly applied in the chase from {\bf a}.l\_root to obtain {\bf a}. In particular
we have ${\bf a} = \rho_n(\ldots(\rho_1({\bf a}.l\_root)))$, where \(\rho({\bf a})\) denotes
the application of a rule \(\rho\) to a fact {\bf a} in the chase. On provenance
we also define the \emph{inclusion relation} \(\subseteq\) as the
set of pairs \(\langle p_i, p_j \rangle\) such that \(p_i\) is an ordered (and possibly coinciding) left-subsequence of
\(p_j\).

The \textbf{ground structure} 
G stores the nodes of the warded forest incrementally built during the chase, grouping them
by the root of the tree in the warded forest to which each node belongs. G is a dictionary of sets
of facts. More precisely, each element G[{\bf a}.w\_root] represents 
the set of facts of the tree rooted in {\bf a}.w\_root in the warded forest.
Based on G, we define the provenance $\lambda = [\rho_1, \ldots, \rho_n]$ for \({\bf a}\) as \emph{non-redundant} 
if every ${\bf a^i} = \rho_i(\ldots(\rho_1({\bf a}.l\_root)))$ for \(i=1,\ldots,n\)
is not isomorphic to any ${\bf a}^j \in G[{\bf a}.w\_root]$.
We define \(\lambda\) as a
\emph{stop-provenance} if it is a \(\subseteq\)-maximal non-redundant provenance.
In more plain words, \(\lambda\) is non-redundant if no intermediate fact in the chase from {\bf a}.l\_root
to {\bf a} is isomorphic to any other fact in the entire tree of {\bf a} of the warded forest; it is
a stop-provenance if every provenance \(\lambda^\prime \supset \lambda\) does not satisfy non-redundancy.
Stop-provenances then represent a maximal root-leaf path in a tree of the lifted
linear forest, i.e., any extension of it is bound to contain a node that is pattern-isomorphic
to another node of the same tree.

The \textbf{summary structure} 
S stores the lifted linear forest, which is incrementally built during the chase.
In particular, it memorizes the complete paths from the roots, which are the patterns of the roots of the underlying linear forest,
to the leaves in the form of stop-provenances.
S is a dictionary of sets of stop-provenances, indexed by common root pattern.
At a given point in time, each element S[$\pi$({\bf a}.l\_root)] of S thus represents the partial tree of the lifted linear forest,
with maximal root-leaf paths.

\begin{algorithm}
\caption{Termination strategy for the chase step.     \label{alg:termination}   }
 \begin{algorithmic}[1]
     \Statex
    \Function{check\_termination}{${\bf a}$}
        \If{a.generating\_rule == \{\textsc{LINEAR} or \textsc{WARDED}\}}
        		\If{$\exists \lambda \in$ S[$\pi$({\bf a}.l\_root)] s.t. $\lambda \subseteq$ {\bf a}.provenance}
        			\State \Return {\bf false}
		 	\Comment{beyond a stop provenance}
		\ElsIf{$\exists\lambda$ $\in$ S[$\pi$({\bf a}.l\_root)] $\mbox{s.t.}$ {\bf a}.provenance $\subset$ $\lambda$}
			\State \Return {\bf true}
			\Comment{within a stop provenance}
		\Else{}
			\Comment{continue exploration}
			\If{$\exists${\bf g} $in$ G[{\bf a}.w\_root] s.t. {\bf a} isomorphic to {\bf g}}
				\State S[$\pi$({\bf a}.l\_root)]={\bf a}.provenance
				\State \Return {\bf false}
				\Comment{isomorphism found}
			\Else{}
				\State G[{\bf a}.w\_root].append({\bf a})
				\State \Return {\bf true}
				\Comment{isomorphism not found}
			\EndIf
        		\EndIf
        \ElsIf{{\bf a} $\notin$ G}
        		\Comment{other non-linear generating rules}
        		\State G[{\bf a}.w\_root].append({\bf a})
		\Comment{and reset provenance}
		\State \Return {\bf true}
	\Else
		\Comment{the new tree is redundant}
		\State \Return {\bf false}
        \EndIf
      \EndFunction
  \end{algorithmic}
\end{algorithm}

\shortRemove{\begin{algorithm}
\caption{A generic chase using the termination strategy. \label{alg:chase}}
 \begin{algorithmic}[1]
     \Statex
      \Function{chase}{$D,\Sigma$}
      \ForAll{$\sigma \in \Sigma$ and  {\bf x} to which $\sigma$ applies}
		\If{{\sc check\_termination}($\sigma({\bf x})$) } 
			\State $D = D \cup \{\sigma({\bf x})\}$
		\EndIf
      \EndFor
      \EndFunction
  \end{algorithmic}
\end{algorithm}
}

\medskip
\noindent
Let us now analyze the dynamic aspects of the algorithm.
We assume that: 1. the rules at hand are harmless warded; 
2. existential quantifications appear only in linear rules. For any set of warded Datalog rules
both conditions can be always achieved: the first by applying the Harmful Joins Elimination Algorithm (Section~\ref{sec:harmless}),
the second with an elementary logic transformation.

Linear and warded rules produce facts {\bf a}, for which, in the base case, isomorphism
checks must be performed. The ground structure allows to restrict this
costly check to the local connected component of the warded forest (line 8-10), featuring
a form of \emph{local detection}. 
If an isomorphic fact is found (line 8), the algorithm stores the stop-provenance
of {\bf a} in such a way that: 1. whenever the same sequence of rules is applied from 
{\bf a}.l\_root, the result can be discarded (enforcement with vertical pruning) without performing any isomorphism check (line 3);
2. whenever a subsequence of rules is applied from {\bf a}.l\_root, no superfluous isomorphism checks
are performed (line 5). 
Moreover, since we want to maximize the reuse of the learnt stop-provenance so as to apply it
to as many facts as possible and independently of the specific ground values (enforcement with horizontal pruning),
the algorithm maps {\bf a}.l\_root into the respective root of the lifted linear forest and stores
the provenance in into S w.r.t. to the pattern \(\pi({\bf a}.l\_root)\) (line 9).
Other, (non-warded) non-linear rules are the roots of new trees (connected components) of the warded forest
(line 14). Since isomorphism
checks are only local to the trees in the warded forest, the current fact provenance can be forgotten for memory reasons. New trees
are generated unless their root is already present in G, which by Theorem~\ref{th:isomorphism_harmless}
would imply that the isomorphism of the entire subtree. As we assume that non-linear rules do not have
existential quantification, the condition can be efficiently checked as set containment of ground facts.

\medskip
\noindent
Our algorithm guarantees chase termination as a consequence of Theorem~\ref{th:isomorphism}. Actually, the set
of distinct symbols that can be present as terms of facts in \emph{subtree}({\bf a}) of the 
warded forest is bounded up to isomorphism of the labelled nulls.
\vsRemove{In particular, along the lines of Guarded Datalog\(^\pm\)~\cite{CaGL12} it would be possible
to conduct a simple combinatorial analysis and establish: 1. an upper bound to the cardinality of such set up
to isomorphism; 2. the maximum theoretical number of
chase steps exceeding which we are bound to re-generate an existing fact (\emph{warded depth}), due to the
finiteness of possible symbols in each subtree.
Instead, }Algorithm~\ref{alg:termination} provides \vsRemove{a much more }\vsAdd{an} effective strategy, by applying
a form of \emph{lazy pattern recognition} that exploits regularities in the warded forest as soon as they are detected.
The roots of the warded trees represent the ``handles'' for detecting such regularities.
Besides, by Theorem~\ref{th:isomorphism_harmless}, we know that relying on such isomorphisms
in the chase graph is correct in the sense that fact isomorphism in the guarantees the isomorphism of the respective subgraphs.
The roots of the lifted linear trees represent the handles for storing the learnt regularities in a generalized way 
and enforcing them afterwards. The generalization of isomorphism of trees to patterns-isomorphism based on roots patterns is correct as a consequence 
of Theorem~\ref{th:linear_forest}.

The most time-demanding operation in Algorithm~\ref{alg:termination} is isomorphism check, which in a na\"ive chase
step implementation would require checking each fact against the entire database. 
We significantly limit the number of isomorphism checks and the cost of the single check.
The lifted linear forest allows to minimize the number of checks, basically avoiding
superfluous controls for pattern-isomorphic facts generated in sub- or super-paths of known ones.
\shortRemove{As an interesting perspective, the technique can be seen as a sophisticated form of dynamic programming where
specific recursive branches are collapsed.
The warded forest limits the cost of the single check, since each fact
is checked only against the other facts in the same tree.}

In terms of memory footprint, S is extremely compact as the lifted linear forest does not contain
the ground values, but only their equivalence classes modulo constant- and variable- isomorphism.
\vsRemove{Facts require only tree-local provenance information and therefore it can be reset (line 17), whenever
non-linear (non-warded) rules are applied. G is actually space-demanding since it stores the ground values.
However, once a warded tree has been completely explored, the ground values for such tree (except for the root)
can be dropped.}

\section{Architecture}
\label{sec:architecture}

\noindent
In Section~\ref{sec:algorithm}, we presented a practically useful algorithm
for supporting reasoning with Warded Datalog\(^\pm\). 
Specifically, Algorithm~\ref{alg:termination} guarantees the termination of
chase-based procedures by controlling the recursion, and is already optimized
in the sense it avoids the typical overheads that chase-based algorithms usually 
need to spend on homomorphism checks.

Yet, the adoption of a traditional chase-based procedure \shortRemove{(such as the na\"ive one in Algorithm~\ref{alg:chase}, or more
advanced ones)}in the presence of large amounts of data has a number of disadvantages, in particular requiring
the entirety of original and generated data to be available as possible inputs for chase steps, untenable for large-scale settings.
In this section, we show how the limitations of the chase are overcome by presenting the specialized
architecture that we devised for the Vadalog system. We show that our termination algorithm plays a fundamental role in it.

\begin{figure} [t!]
\centering
\hspace{-4mm}
    \includegraphics[scale=0.35]{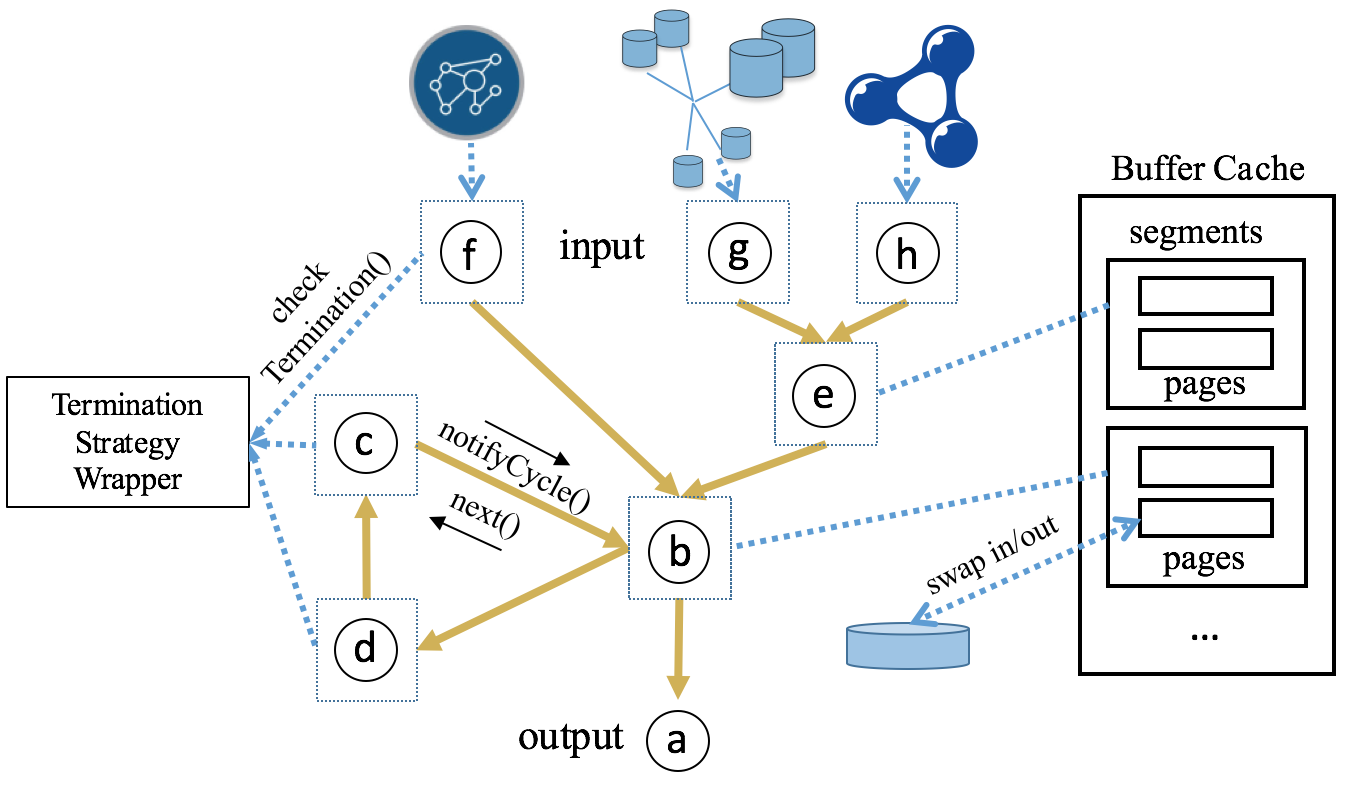}
\caption{The pipeline architecture of the Vadalog system.}
\label{fig:pipeline}
\vspace{-3mm}
\end{figure}

{\bf Pipeline architecture}. One of the core design choices in the implementation of the Vadalog system is the
use of the \emph{pipe and filters} architectural style, where the set of logic rules and the queries are
together compiled into an active \emph{pipeline} that reads the data from the input sources, performs the needed transformations, and
produces the desired output as a result. This is actually done in four steps. 1. A  \emph{logic optimizer} applies
several logic transformations to the original set of rules, consisting in elementary (e.g., multiple head elimination, elimination of redundancies, etc.)
and complex (e.g., elimination of the harmful joins) rewritings of the rules. 2. A \emph{logic compiler}
reads the object representation of the rules received from the \emph{rule parser} and transforms 
it into a \emph{reasoning access plan}. The reasoning access plan is a logic pipeline, where each rule of the program corresponds to a filter (i.e., a node) in the pipeline
and there is a pipe (i.e., an edge) from one filter \(a\) to a filter \(b\), if rule \(a\) has in the body an atom that unifies with the head of \(a\). 3. The \emph{execution
optimizer} transforms the reasoning access plans, performing basic optimizations, such as rearranging the join order, pushing selections
and projections as close as possible to input nodes and so on. 4. A \emph{query compiler} turns the reasoning access plan into a 
\emph{reasoning query plan}, an active pipeline (shown in Figure~\ref{fig:pipeline}), capable to perform the needed transformations (e.g.,
projections, selections, joins, application of functions, creation of new values, aggregations, etc.) and producing the output data.

{\bf Execution model}. In {\sc Vadalog}, some rules are marked as ``input'' (by specific annotations) as
they represent external data sources and therefore are mapped to source filters in the query plan; 
vice versa, terminal nodes are marked as ``output'' and
represent sink filters in the pipeline. The reasoning process is then realized as a data stream along the pipe\-line, implemented
with a \emph{pull (query-driven) approach}. We implement a generalization of the \emph{volcano iterator model}~\cite{GrMc93} used in DBMSs,
extending it to the entire pipeline. Each filter pulls the required input from the respective sources, which in turn,
pull from their sources down to the initial data sources. The initial data sources, use \emph{record managers}, specific
components that act as adapters towards external sources, turning input streaming data into facts. Therefore the entire process is triggered 
and driven by the sink nodes, which issue \texttt{open()},  \texttt{next()}, \texttt{close()}
messages to their predecessors, which receive and propagate the messages back down to the source nodes. 
The \texttt{next()} primitive returns a Boolean indication of the presence of facts for a specific requestor. 
The facts are then accessed with specific typed \texttt{get} primitives. 
The behavior of the \texttt{next()} primitive is driven by the direct availability of facts in the invoked filter as 
well as further logics that control termination according to Algorithm~\ref{alg:termination}, as we will see.

Since for each filter, multiple parent filters may be available, our execution model adopts a \emph{round-robin strategy}, that is,
it tries to pull data from all the available parent filters, in a predefined order. We empirically verified that such strategy
guarantees a good balance of the workload among the filters and a uniform propagation of the facts
from the sources to the sinks. From another perspective, the round-robin strategy sustains a breadth-first application of the rules,
where a rule is re-applied, only when all the others have been applied in turn (successfully or not), and in the same order.
In the example in Figure~\ref{fig:pipeline}, the output filter \(a\) sends
a \texttt{next()} message to \(b\), which is propagated to \(c\), \(f\) and \(e\).

There are two factors that make the stream-based processing definitely non-trivial: not surprisingly, the presence
of typically ``blocking'' operations (in particular, the join and the aggregations) and, more important to our case, 
the possible cycles in the pipeline, which can be induced by the recursion in the rules and can lead to non-termination of the streaming process.

\shortRemove{{\bf Slot machine join}. Let us introduce our join technique for unary predicates $A_1(x),...,A_n(x)$,
since it can be easily extended to arbitrary arity, as the system actually supports. 
The slot machine join is an extension to the \emph{indexed nested loop join}, 
enhanced with \emph{dynamic in-memory indexing}.
For each predicate \(A_i\) we have an \emph{iterator}, providing two primitives: \texttt{next()},
which proceeds to the next value (fact for the predicate \(A_i\)), if any, and
\texttt{get}$(c)$, returning the fact \(A_i(x)\), with \(x=c\), if existing.
Initially the iterators are located at the first fact for each predicate.
Then, for all the predicates \(A_i\), with \(0 \le i \le n\),
for each fact $A_i(c)$ returned by the application of \texttt{next()} on $A_i$, the iterator on 
$A_{i+1}$ is consumed with an \emph{index-based access} with the \texttt{get}$(c)$ 
primitive, or actually with a \emph{full scan}, performing \texttt{next()} until
a matching fact is met (if any). While this is of no particular novelty,
what distinguishes our algorithm from usual nested loop join is that there is
no persistent pre-calculated index; instead, while the various iterators 
are consumed by a full scan, an index is 
dynamically built in memory for each of them (\emph{dynamic indexing}). 
These dynamic indexes can be used even when still incomplete.
We use an optimistic approach: first we try a \texttt{get()} call
on the index and in case of failure (\emph{index miss}), if the index is incomplete, an actual full scan
is performed. The technique is essential in the presence of a pervasive
recursion, where usual persistent indexing is almost inapplicable.
As we adopt \emph{hash structures}, the cost of our join 
tends to the number of facts for predicate \(A_1\).}

\vsRemove{{\bf Non-blocking monotonic aggregations}. }\longpaper{As we will see in Section~\ref{sec:extensions},}
we support non-blocking aggregation as a specific feature of
the {\sc vadalog} language, which avoids the presence of blocking nodes in this case.

{\bf Cycle management}.\shortRemove{We deal with two kinds of cycles: \emph{runtime invocation cycles} and
\emph{non-terminating sequences}.
Runtime cycles are recursive cases in the invocation of the \texttt{next()} primitive, like in
the sequence $a \leftarrow b \leftarrow c \leftarrow d \leftarrow b$ in Figure~\ref{fig:pipeline}.
If \(b\) cannot fulfill the recursion with any fact 
from other recursive or base cases, we are still not allowed to interrupt the iteration, as
the required facts for \(b\) may derive from other recursive cases to be explored first.
We distinguish the absence of facts in cyclic cases (\emph{cyclic miss}) from the actual absence of
further facts (\emph{real miss}). While the first may be temporary and due to the non-deterministic order
of recursive rule execution, the second is permanent and denotes that a filter does not have --- and will not have ---
facts, thus justifying a negative reply from \texttt{next()}. In case of cyclic misses, the invoked filter notifies the caller
with a \texttt{notifyCycle()} primitive, that flows along the pipeline from called filters to caller. In the example, the original invoker \(b\)
is eventually informed that the request could not be satisfied due to the presence of a cycle; at that point, \(b\) will explore
other predecessors until either one suitable record is found or none is found. 
If all the predecessors issue a \texttt{notifyCycle()},
it indicates that the recursion cannot be satisfied and the runtime cycle is actually turned into a negative answer. 
\longpaper{Notice that
the adoption of the round-robin policy is essential, as it allows a fair and balanced exploration of all the possible
recursive and base cases.}
}
Non-terminating sequences directly derive from the presence of recursion (and hence cycles): in absence of further checks in the sink filters,
a recursive pipeline may generate infinite facts. To cope with this, we feature \emph{termination strategy wrappers},
components that implement Algorithm~\ref{alg:termination} and work as follows. When a filter receives
a \texttt{next()} call, it agnostically pre-loads a fact $A(c)$, in the sense that it issues all the necessary \texttt{next()} primitives
to its predecessors, handles the runtime cycles, so that \(A(c)\), if available, is stored in the filter data structure
to be possibly consumed. The filter issues a \texttt{checkTermination(}A(c)\texttt{)} message to its local termination wrapper,
which applies Algorithm~\ref{alg:termination}. The termination strategy wrappers
also manage the \emph{fact}, \emph{ground} and \emph{summary} structures of Section~\ref{sec:algorithm}.
Then, if the termination check is negative, $A(c)$ is discarded 
as it would lead to non-termination.

\shortRemove{{\bf Memory management}. The {\sc Vadalog} system adopts a full in-memory processing architecture: ideally the resulting facts are
materialized only at the end of the elaboration and if desired. For performance reasons 
the intermediate facts produced by the single filters are stored in a  \emph{buffer cache}.
We adopt a \emph{fragmented buffer management} scheme, where
each filter in the pipeline is wrapped by a cache component which maintains intermediate results. In another
perspective, we put a specific \emph{buffer segment} at the disposal of each filter. Within each buffer segment, we then handle \emph{buffer pagination}
with eviction and swap to secondary storage with usual techniques (e.g., LRU, LFU, etc.). Each specific buffer segment
is mapped into a memory area of the overall buffer cache, which is the collection of all the segments.
We primarily use the buffer cache as proxies for the \texttt{next()} calls, which means, to inhibit actual calls in case of \emph{cache hits}.
Besides, we use the buffer cache to provide buffer-aware implementations of our algorithms. 
For example, the dynamic indexes of the slot machine join and
the data structures of the termination strategy wrappers reside in the buffer cache
and so inherit the pagination (and memory eviction) support.}

\longpaper{\section{Advanced Vadalog Features}
\label{sec:extensions}

\noindent
Although the Warded Datalog\(^\pm\) offers a solid framework for the core of our
{\sc Vadalog} Reasoner, a full set of features of practical utility are needed to support
real-world applications. While for the core, tractability is guaranteed, in general the adoption of
particular combinations of the advanced features may lead to highly complex and in some
case not even terminating programs.

\medskip 
\noindent
{\bf Data Types}. {\sc vadalog} terms (variables and constants) are indeed typed. We support basic data types such as
integer, float, string, date, Boolean as well as composite data types such as sets, lists and dictionaries.
We will not specifically refer to data types any longer in the paper, though
their adoption will be implicit.

{\bf Expressions.} We support expressions in the body of rules with two main goals: 1. as the left-hand side (LHS)
of a condition, which is the comparison (\(>,<,\ge,\le,\neq\)) of a body variable with the expression; 2.
as the LHS of an assignment, which allows to specifically define a value for an existentially quantified
variable in the head. We define expressions recursively as follows: a term is an expression;
a combination of expressions is an expression. Actually expressions can be
combined by means of a selection of type-related operators, which we provide, such as:
algebraic (+, -, *, /, exponentiation, etc.), string (startsWith, substring, indexOf, etc.), date, boolean operations and so forth. Clearly expressions are typed and type conversion operators are available as well.

{\bf Skolem Functions}. Values for existentially quantified variables can be conveniently calculated with Skolem functions,
which allow to control the identify of the labelled nulls. A {\sc vadalog} Skolem function
is an expression of the form \(\#f(\bar x)\), where \(f\) is any function name, and \(\bar x\) is
an n-uple of terms (\(n\) is the arity of \(f\)). Our Skolem functions are \emph{deterministic} (they
produce the same labeled null for repeated invocations), they are \emph{injective} and 
\emph{range disjoint}.

{\bf Monotonic Aggregation}. Real-world applications require support for aggregate functions, namely
\emph{sum}, \emph{min}, \emph{max}, \emph{count} as well as SQL-like grouping constructs. 
The {\sc vadalog} language supports these by bringing the notion of
\emph{monotonic aggrega\-tion}~\cite{ShYZ15} in the Datalog\(^\pm\) context in such a way
that monotonicity w.r.t.\ set containment is preserved.

Intuitively, we provide aggregate functions as stateful record-level operators, which
memorize the current aggregate value. Subsequent invocations
of a function then yield updated values for the aggregate so as that the ``final value''
is the actually desired aggregate value.
The production of intermediate aggregates is unusual, especially w.r.t. SQL aggregates. Moreover,
the value of intermediate aggregations is non-deterministic since it depends on the
specific chase sequence.
Nevertheless, the monotonicity of the aggregate functions
guarantees that the ``final value'' is unambiguously defined (i.e., it is the minimum or the maximum)
and the approach does not enforce a specific order of the chase steps, harmlessly introducing
aggregations in the usual procedure. We support group-by arguments as usual as well as 
windowing functionality, which is the possibility to define sub-groups in each group-by group
and choose only the minimum (maximum) value within that sub-group.

Formally, a rule with an aggregation is a first-order sentence of the form
\[
\forall \bar x (\varphi(\bar x), z = {\rm maggr}(x, \langle \bar c \rangle) \ \rightarrow\ \psi(\bar g, z))
\]
where {\rm maggr} is the name of an aggregation function, \(x \in \bar x\), and
\(\bar g \subseteq \bar x\) is a \(n\)-uple of \emph{group-by
arguments}, \(\bar c \subseteq \bar x\) (with \(\bar c \cap \bar g = \emptyset\)) is a \(m\)-uple of \emph{contributor variables} (or simply \emph{contributors}) and \(z\) is a \emph{monotonic aggregate} (an existentially quantified variable whose value is computed by the aggregation).

Let us consider tuples \(\langle \bar g, \bar c, x_i\rangle\) of \(\varphi(\bar x)\)
and let \(X_{\bar g}\) be each of the multi-sets of tuples of \(\varphi(\bar x)\) selected by
a specific n-uple of \(\bar g\). An aggregate rule maps each multi-set \(X_{\bar g}\)
into an output set \(Z_{\bar g}\) of tuples \(\langle \bar g, z_i\rangle\) of \(\psi(\bar g, z)\) computed as follows.
For a monotonically decreasing (increasing) aggregation function, 
for each \(\langle \bar g, \bar c, x_i\rangle \in X_g\) we have a corresponding 
tuple \(\langle \bar g, z_i\rangle \in Z_{\bar g}\), where the value for \(z_i\)
is computed as \(z_i = {\rm maggr}(\bar c, x_i)\). These aggregate functions memorize the most
recently computed aggregate for the group and return an updated value at each invocation in such a way that
for each value of \(\bar c\), the minimum (maximum) value of \(x_i\) is considered in the
current aggregate. Let us clarify the behavior with the following example

\begin{example}
\label{ex:aggregates}
\begin{eqnarray*}
{\rm P}(1,2,5).~{\rm P}(1,2,3).~{\rm P}(1,3,7).~{\rm P}(2,4,2).~{\rm P}(2,4,3).~{\rm P}(2,5,1). \\
{\rm P}(x,y,w), j={\rm msum}(w, \langle y \rangle) \rightarrow \underline{\rm Q}(x,j).
\end{eqnarray*}
Here we are aggregating over \(w\) and \(x\) is the group-by argument;
we have two groups in the result, for \(x=1\) and for \(x=2\).
Let us suppose the chase considers the facts in the order: for \(P(1,2,3)\), it generates
\({\rm Q}(1,5)\); then, since \({\rm P}(1,2,3)\) has the same contributor, \(2\), as \({\rm P}(1,2,3)\),
the chase considers \(\mbox{max}(5,3)\) and returns again \({\rm Q}(1,5)\).
Then \({\rm P}(1,3,7)\) has a different contributor, \(3\), hence the sum is actually produced and we
obtain \({\rm Q}(1,12)\). For the second group we obtain in the order: \({\rm Q}(2,2)\), \({\rm Q}(2,3)\) (as for contributor
\(4\) we take \(\mbox{max}(2,3)\), and \({\rm Q}(2,4)\).
\end{example}

\noindent
Example~\ref{ex:aggregates} shows a particular chase sequence; clearly any
sequence is allowed and the monotonicity of the functions guarantee that the minimum (maximum in the example)
value for each group corresponds to the final aggregate. Another instance
of aggregations is in Example~\ref{ex:running}, where we compute the ownership quota
of \(x\) on \(z\), by summing the quotas of each distinct contributor \(y\).

Monotonic aggregations are a generalization of the usual aggregations such as the ones in SQL,
which correspond to the cases when \(\bar c = \emptyset\), i.e., all the tuples in each group contribute to the aggregate.
The role of \(\bar c\) is that of \emph{sub-grouping} or \emph{windowing} in \emph{analytic} extensions to SQL.\footnote{See for example \url{https://docs.oracle.com/cloud/latest/db112/SQLRF/functions004.htm#SQLRF06174}}

The following constraints to the adoption of aggregations are assumed: 1. the group-by arguments
and the contributors must be non-null values; 2. when a position \(\pi\) is calculated as
an aggregation, \(\pi\) must be always calculated with the same aggregation; 

As a further extension, in {\sc vadalog}, actually \(z\) can be calculated as an arbitrary
expression, which may include the usual operators, as explained, as well as aggregations. 
Similarly, argument \(x\) of the aggregate function can in turn be a numeric expression. 

{\bf Annotations}. The {\sc vadalog} language features a behavior injection mechanism,
based on special ``\texttt{@}''-prefixed facts, for various purposes, including:
binding predicates to data sources or to queries, with specific input/output patterns;
procedural post-processing.

\emph{Dynamic Source Binding}: The \texttt{@bind} and \texttt{@qbind} annotations bind \texttt{@input}- or \texttt{@output}-annotated 
predicates to sources or queries so that the Reasoner will derive facts from the external sources, in a form of 
factual interface integration. A \texttt{@mapping} annotation can be used, when needed, to harmonize the named perspective
of many sources with the unnamed positional one of {\sc vadalog}.

\emph{Post-processing Directives}: They allow to apply final post-pro\-ces\-sing steps to the
\texttt{@output}-annotated predicates, such as: sorting by specific attributes; computing
aggregates over specific attributes according to the usual SQL aggregate semantics; 
dropping facts with labelled nulls, returning in this way a \emph{certain answer}~\cite{FKMP03}.
}

More discussion on the vision of the Vadalog language in the area of data  wrangling and necessary extensions in this context can be found in \cite{amw/FurcheGNS16}. A discussion on data extraction and the OXPath language that is used as an advanced feature in the Vadalog language can be found in a paper in collaboration with dblp \cite{jcdl/MichelsFLSS17}.

\section{Experimental Evaluation}
\label{sec:experiments}

\figurehere

\vsRemove{\begin{figure*}[t!]
\begin{scriptsize}
\begin{center}
\begin{tabular}{|c|c|c|c|c|c|c|c|c|c|}
\hline
\pbox{1.5cm}{{\bf set of rules}} &
\pbox{1.55cm}{{\bf L rules}} & 
\pbox{1.55cm}{{\bf $\Bowtie$ rules}} & 
\pbox{1.3cm}{{\bf L recursive}} & 
\pbox{1.3cm}{{\bf $\Bowtie$ recursive}} &  
\pbox{1.45cm}{{\bf $\exists$ rules}} & 
\pbox{1.45cm}{{\bf hrml $\Bowtie$ hrmf }} & 
\pbox{1.45cm}{\vspace{1mm}{\bf hrml $\Bowtie$ hrml } \\with ward\vspace{1mm}} & 
\pbox{1.45cm}{{\bf hrml $\Bowtie$ hrml } \\w/o ward} & 
\pbox{1.45cm}{{\bf hrmf $\Bowtie$ hrmf }}
\\
\hline
\hline
\emph{synthA} & 90 & 10 & 27 & 3 & 20 & 5 & 4 & 1 & 0  \\
\hline
\emph{synthB} & 10 & 90 & 3 & 27 & 20 & 45 & 40 & 5 & 0 \\
\hline
\emph{synthC} & 30 & 70 & 9 & 20 & 40 & 25 & 20 & 5 & 20 \\
\hline
\emph{synthD} & 30 & 70 & 9 & 20 & 22 & 10 & 9 & 1 & 50 \\
\hline
\emph{synthE} & 30 & 70 & 15 & 40 & 20 & 35 & 29 & 1 & 5 \\
\hline
\emph{synthF} & 30 & 70 & 25 & 20 & 50 & 35 & 29 & 1 & 5 \\
\hline
\emph{synthG} & 30 & 70 & 9 & 21 & 30 & 0 & 10 & 60 & 0  \\
\hline
\emph{synthH} & 30 & 70 & 9 & 21 & 30 & 0 & 60 & 10 & 0  \\
\hline

\end{tabular}
\end{center}
\end{scriptsize}
\caption{Details of the scenarios generated with {\sc iWarded}.}
\label{tab:synthetic_programs}
\end{figure*}}

\noindent
In this section we evaluate  the performance of the Vadalog system 
on two synthetic (Sections~\ref{sec:synth} and~\ref{sec:iBench}) 
and two real-world scenarios (Sections~\ref{sec:dbpedia} and~\ref{sec:industrial_validation}), all involving
non-trivial war\-ded Datalog$^\pm$ rules (with many existentials, harmful joins, null propagation, etc.). 
We thoroughly validate our theoretical results and the architecture on such cases and show that 
our reasoner exhibits very good scalability and outperforms existing systems.

As a supplementary experimental contribution (Section~\ref{sec:chasebench}), we then show
that the Vadalog system is also a best-in-class general purpose 
chase/query answering/data integration system, by comparing
its results with top-performing systems on sets of rules where the typical characteristics of
wardedness cannot be exploited.
\cbstart
\new{We highlight the advantage of our approach over pure isomorphism checks
with specific experiments (Section~\ref{sec:homomorphism}) and
evaluate scalability along many
different dimensions (Section~\ref{sec:dbsize}).}
\cbend

\medskip
\indent \textbf{Test setup.} 
The reasoner was used ``as a library'' and invoked from
specific Java test classes for end-to-end, i.e., storage to storage executions of the reasoning.
For the storage, we adopted simple CSV archives to precisely highlight the
performance of the reasoner, independently of any optimized back-end
physical structures, such as indexes, and so on. Clearly, in some
cases our tasks could benefit from back-end rewriting (e.g., performing joins or aggregations in the DBMS\longpaper{, as
also allowed by our \texttt{\@qbind} directive illustrated in Section~\ref{sec:extensions}}),
yet here we intentionally execute every operation purely inside of our system.
We also used local installations of PostgreSQL 9.6, Oracle 11g, MySQL 5 and Neo4J 3.2.

\medskip
\indent \textbf{Hardware configuration.} We ran the tests on a Linux server with 8 Xeon v3 cores
running at 2.4 GHz and with 16GB of RAM.

\subsection{iWarded: Synthetic Warded Scenarios}
\label{sec:synth}

\noindent
In this section we investigate the impact of specific properties of the warded rules
on the performance of reasoning tasks in the Vadalog system. To this end, we developed {\sc iWarded}, a flexible
generator of warded rules that allows to control the internals related
to Warded Datalog$^\pm$ such as the number of linear and non-linear rules, the presence of harmful or harmless
joins, recursion, etc. 

{\bf Description of the scenarios}. \vsRemove{In Figure~\ref{tab:synthetic_programs} we summarize our scenarios
and the respective parameters. In particular, for each set of rules
we set the number of: linear rules (\emph{L rules}); recursive linear rules (\emph{L recursive}); non-linear rules ($\Bowtie$ \emph{rules}); recursive non-linear rules ($\Bowtie$ recursive);
linear and non-linear rules that have existential quantification ($\exists$ \emph{rules});
joins between harmless variables where one involved atom is a ward (\emph{hrml} $\Bowtie$ \emph{hrml with ward}); joins between harmless variables where none of the atoms is a ward (\emph{hrml} $\Bowtie$ \emph{hrml w/o ward}); joins between harmful variables (\emph{hrmf} $\Bowtie$ \emph{hrmf}).

}We built eight scenarios, all with 100 rules and the same set of (multi-)queries that
activates all the rules.
\emph{SynthA} has a prevalence of linear rules and 20\% of the total rules
have existential quantification; 30\% of linear and non-linear rules are recursive,
and the joins are equally distributed between harmless-harmful and 
harmless-harmless, for which we have a prevalence of joins involving wards.
\emph{SynthB} is specular to  \emph{SynthA}, but has a prevalence of non-linear rules;
all the other proportions are respected accordingly. 
\emph{SynthC} (as well as all the next settings) has 
a 30\%-70\% balance between linear and non-linear rules, which we posed
to highlight the impact of the (various kinds) of joins on the performance. 
In this setting, harmful-harmful join are present.
\emph{SynthD} stresses the presence of such joins even more.
\emph{SynthE} studies the impact of a strong presence of recursion of the non-linear rules. 
\emph{SynthF} deals with a high recursion incidence on linear ones. 
\emph{SynthG} has basically the same characteristics as \emph{SynthC} and
\emph{SynthD}, but with a prevalence of harmless rules that do not involve
wards, therefore its characteristics resemble those of a pure Datalog program.r
\emph{SynthH} emphasizes the joins on wards.

{\bf Results}. The results are reported in Figure~\ref{fig:experiments}(a).
The  Vadalog system shows the best performance in scenarios \emph{SynthB}
and \emph{SynthH}, for which the execution times are under \old{20}\new{10} seconds. The prevalence of join rules, with particularly high values for harmless \specialbreak joins with wards allows to exploit the wardedness at best. Warded
joins produce particularly deep warded forests, making the isomorphism check for each
generated fact particularly effective: the deeper the warded forests are, the more
likely it is that an isomorphic fact is met within the same structure during the isomorphism check in
the ground structure. Therefore the summary structure is updated more frequently, maximizing the pattern learning and letting the algorithm converge sooner. 
\vsRemove{Observe that the 45
harmless-harmful joins in \emph{SynthB} do not overly affect the performance and only a \old{0.2}\new{1} second
difference is noticeable. Indeed, certainly the presence of such joins tends to break
trees in the warded forest into multiple ones, hence hampering the beneficial effects of the deep forests; 
however the joins actually produce values only for ground inputs and their effect on the proliferation
of new trees (and new patterns) is therefore limited.} Then, scenario \emph{SynthC} acts as a baseline,
with an average number of rules and balanced join types. Although scenario \emph{SynthD}  has less than
half the harmless joins with ward of \emph{SynthC}, it requires 7 seconds more due to
a higher number of harmful joins.
\vsRemove{The harmful joins elimination algorithm reduces these joins into
harmless-harmless without ward, i.e., when possible, harmful joins are linearized. This causes
some fragmentation of the warded forest, as we have seen for \emph{SynthB}.}
Scenario \emph{SynthG}, has 60 harmless joins without wards, i.e., it resembles
the result of the harmful join elimination applied to \emph{SynthD} and therefore
shows almost the same times, though slightly affected
by the presence of more existentials that give rise to more patterns. \emph{SynthG} is indeed
very important, since it pinpoints the behavior of the Vadalog system in the presence
of generic Datalog programs, where the properties of warded rules cannot be exploited.
The performance is indeed good, as is also confirmed in Section~\ref{sec:chasebench}
by the comparison with chase-based systems, though typical optimizations of Datalog
(foreseen as a future optimization) will bring improvements. Finally, scenarios
\emph{SynthE} and \emph{SynthF} show that the impact of recursion on performance
is significant, with times of about 40 and 65 seconds, respectively. 
\vsRemove{Interestingly, extreme linear recursion appears very impactful. 
We explain it with a multiplicative effect that the injection of many nulls in the linear rules has
in combination with the harmful joins: there is a proliferation of the tree structures
in the warded forest, undermining the beneficial effects of wardedness. This is confirmed
by a comparison with \emph{SynthA}, which only requires 32 seconds. In this latter case, 
the linear recursion is stronger (and linear rules are even 90); nevertheless, the impact on
times is limited as the recursion only produces a lengthening of the trees, since the
absence of harmful joins prevents the fragmentation of the forest which, instead, we
observed in \emph{SynthF}.}

The memory footprint is confirmed to be limited. All the scenarios require 
less than 400MB of memory. Interestingly, in the top-performing
scenarios, all the required memory is immediately allocated. This confirms that
deep warded forests cause a rapid growth of ground and summary structures.

\subsection{iBench}
\label{sec:iBench}
\vsRemove{\noindent{\sc iBench}~\cite{AGCM15} is a popular tool to generate \emph{data integration scenarios}, i.e.,
set of dependencies (and thus of logic rules) that are very large and
complex, yet can resemble realistic settings by offering different tunable characteristics.
In general, {\sc iBench} generates second order tgds (\emph{tuple-generating dependencies}) but,
indeed, we considered two popular interesting first-order integration 
scenarios, namely \emph{STB-128} and \emph{ONT-256}, which we translated from 
the language of tgds into {\sc Vadalog}.
These scenarios are particularly relevant for our purposes. First, they happen to produce
sets of rules that are non-trivially warded, in the sense that there are many existentials,
labelled nulls are often propagated to the heads and also involved in many joins
of different kinds. Besides, the sets of rules are highly recursive. Then, as 
\emph{STB-128} and \emph{ONT-256} have been included in a recent benchmark, {\sc ChaseBench}~\cite{BKMM17},
for chase-based systems, we have the opportunity to compare
the performance of the {\sc Vadalog} system 
on non-trivially warded programs with the top-performing chase-based systems, namely
RDFox~\cite{MNPH14}, {\sc LLunatic}~\cite{GMPS14}, DLV~\cite{LPFE06}, Graal~\cite{BLMR15} and PDQ~\cite{BeLT15}.
}

{\bf Description of the scenarios}. 
\emph{STB-128}: It is a set of about 250 warded rules, 25\% of which
contain existentials; there are also 15 cases
of harmful joins and 30 cases of propagation of labeled nulls with warded rules.\vsRemove{Rules are particularly complex, involve multiple joins and pervasive recursion (originally in the form
of t-tgds (\emph{target tuple-generating dependencies}). There are 112 distinct atoms,
and the source instance contains 1000 facts for each.} The expected target instance
contains 800k facts, with 20\% of labelled nulls.
On this scenario we ran 16 different queries and averaged the
response times. Queries involve on average 5 joins, harmful in 8
cases.

\emph{ONT-256}: It is a set of 789 warded rules, 35\% of which contain existentials; there are
$295$ cases of harmful joins and more than $300$ propagations of labelled nulls. Rules
are even more complex than \emph{STB-128}, and contain multiple joins as well as pervasive recursion.
\vsRemove{There are 220 distinct predicates, with 1000 facts for each in the source instance.}The expected
target instance contains \(\sim\)2 million facts, with an incidence of 50\% of labelled nulls.
On this scenario we ran 11 different queries and averaged the response times. Queries involve
an average of 5 joins, which in 5 cases are harmful.

\textbf{Results}. In Figure~\ref{fig:experiments}(b) we depict the times obtained
for the Vadalog system as well as those reported in {\sc ChaseBench}
for the other mentioned systems in a comparable hardware/software environment.
The Vadalog system outperforms all  systems in both the scenarios: with 6.59 seconds
for \emph{STB-128} and 51.579 seconds for \emph{ONT-256}, it is on average 3 times
faster than the best performing chase-based system, RDFox and 7 times faster than {\sc LLunatic}.
In particular, the three best performing systems, RDFox, {\sc LLunatic} and DLV reported 
28, 52 and 48 seconds for  \emph{STB-128} and 83, 367, and 118 seconds for \emph{ONT-256}.

\subsection{DBpedia}
\label{sec:dbpedia}

\noindent
We built four reasoning tasks based on real datasets about companies and persons
extracted from DBpedia~\cite{dbpedia}.

{\bf Description of the scenarios}. 
For the DBpedia entity \texttt{Company}, we considered the company name,
the \texttt{dbo:parentCompany} property, holding all the controlled companies, 
and \texttt{dbo:keyPerson}, holding all the persons relevant to the company.
We mapped parent companies into
facts {\rm Control}$(x,y)$, if company \(x\) controls
\(y\), and the key persons into facts {\rm KeyPerson}$(x,p)$, if a company $x$ has a key person $p$.
DBpedia publishes \(\sim\)67K companies and \(\sim\)1.5M persons.

\emph{PSC}: The PSC (\emph{persons with significant control}) are 
the set of the persons that directly or indirectly have some control on a company.
The goal of this reasoning setting is finding all the PSC for all the companies of DBpedia.
\begin{example}
\label{ex:psc}
${\rm KeyPerson}(x,p), {\rm Person}(p) \rightarrow {\rm PSC}(x,p)$\\
${\rm Control}(y,x), {\rm PSC}(y,p) \rightarrow {\rm PSC}(x,p).$\\
\end{example}

More precisely, we say that 
a person \(p\) is a PSC for a company \(x\) if either \(p\) is a key person for \(x\), or
\(x\) is controlled by a company \(y\) and \(p\) is a PSC for \(y\).
We ran the scenario, reported in Example~\ref{ex:psc},
for all the available companies (67K) and for subsets of
1K, 10K, 100K, 1M and 1.5M of the available persons. We also executed
the same scenario with indexed tables in PostgreSQL, MySQL, Oracle and graphs in Neo4J,
translating the set of rules of Example~\ref{ex:psc} into recursive SQL and 
Cypher, respectively.

\emph{AllPSC}: This scenario \vsRemove{(Example~\ref{ex:allPsc}) }is a variation of the previous one, where we actually want to
group in a single set all the PSC for a company.
\vsRemove{Also in this case we considered 67K companies
and subsets of 1K, 10K, 100K, 1M and 1.5M of the available persons.

\begin{example}
\label{ex:allPsc}
$ {\rm KeyPers}(x,p), {\rm Pers}(p), j = {\rm munion}(p) \rightarrow {\rm PSC}(x,j)$\\
${\rm Control}(y,x), {\rm PSC}(y,s), j = {\rm munion}(p) \rightarrow {\rm PSC}(x,j).$\\
\end{example}
\vspace{-4mm}
}

Towards the next scenarios,
we say that if companies \(x\) and \(y\) share more than 
a number $N$ of PSC, then there is a \emph{strong link} between them.
Every company \(x\) has at least one PSC, 
which can be either a {\rm KeyPerson} for the company or another, unknown person.
\longpaper{Also here the notion of PSC propagates along the company control relationship.}
Based on the set of rules in Example~\ref{ex:strong_links}
\longpaper{, which represent the illustrated domain,}
we built two scenarios.

\begin{example}
\label{ex:strong_links}

${\rm KeyPerson}(x,p) \rightarrow {\rm PSC}(x,p)$\\
${\rm Company}(x) \rightarrow \exists p~{\rm PSC}(x,p)$\\
${\rm Control}(y,x), {\rm PSC}(y,p) \rightarrow {\rm PSC}(x,p)$\\
${\rm PSC}(x,p), {\rm PSC}(y,p), x>y, w \ge N,\\ \indent\indent\indent  w = {\rm mcount}(p) \rightarrow {\rm StrongLink}(x,y,w)$\\
\end{example}
\vspace{-4mm}

\emph{SpecStrongLinks}: We want to obtain all the strong links of
a specific company, \texttt{Premier\_Foods}, with $N=1$ (which means
that if a company shares one PSC with  \texttt{Premier\_Foods}, it
is a strong link). We considered 1K, 10K, 25K, 50K and 67K companies.

\emph{AllStrongLinks}: We want all the strong links for all the possible pairs of companies.
In order to contain the explosion of the possible strong links, we set $N=3$.
\vsRemove{Also in this case we considered 1K, 10K, 25K, 50K and 67K companies.}
 
{\bf Results}. The results for scenarios \emph{PSC} and \emph{AllPSC}, 
graphed in Figure~\ref{fig:experiments}(c), are particularly meaningful
since they expose the behavior of the Vadalog system in the transitive closure (or reachability) setting. 
The Vadalog system shows very good performance, with linear growth of times and 
absolute values under 100 seconds. Since the facts for PSC are recursively generated,
the good performance confirms the effectiveness of dynamic indexing
in combination with our implementation of the nested loop join. 
In the figure, the lines for \emph{PSC} and \emph{AllPSC}
almost coincide, proving that monotonic aggregations are effective and do not cause overhead.
Interestingly, our reasoner resulted 
two times faster than Neo4J, a best-in-class specialized graph processing system, even for such a
typical graph-oriented task. 
In the figure, we also show that we outperform relational systems, with
PostgreSQL, MySQL and Oracle reporting a 6-times worse performance.

The results for the scenarios \emph{SpecStrongLinks} and \emph{AllStrongLinks} are
depicted in Figure~\ref{fig:experiments}(d). 
In spite of their apparent simplicity, the settings are extremely challenging: 
the transitive closure of PSC along {\rm Control} produces a
very high number of PSC for each company. Moreover,
even when a company does not have any defined {\rm PSC} 
(because there are no direct or inherited key persons), the existential quantification introduces 
an artificial one, which is also propagated with the {\rm Control} relationship to the company subsidiaries. 
Therefore, {\rm PSC} tends to have high cardinality, e.g., 38K facts for 67K companies, which
results in a theoretical number of \(\sim\)700 millions of possible comparisons to be performed in rule 4.
Since our reasoner showed very good performance in the computation of a transitive closure, 
the steep growth curve for \emph{AllStrongLinks}
is explained by the huge size of the output (more than 10 million links), which in turn derives
from very long control chains and moderate density of the underlying graph.
Scenario \emph{SpecStrongLinks} shows almost constant time, always
under 40 seconds, which validates our indexing and query planning mechanisms.

\subsection{Industrial Validation}
\label{sec:industrial_validation}

\noindent
As an industrial validation, we evaluated 
the system against \vsAdd{a variation of Example~\ref{ex:running2}} 
in real-world cases, solving the control problem for the European financial companies,
and in synthetic settings, with generated graphs for large-scale evaluation.

{\bf Description of the scenarios}.
We considered five real ownership graphs, 
built as subsets of the European graph of financial companies, with
10, 100, 1K, 10K and 50K companies and up to \(42K\) edges.
We chose two real-world scenarios:
\emph{AllReal}, where we ask for the control relationship
between all companies and report the reasoning time; \emph{QueryReal},
where we make 10 separate queries for specific pairs of companies, reporting the
average query time.

For the synthetic settings, we built control graphs as 
\emph{scale-free networks}, i.e., graphs whose degree distribution asymptotically
follows a particular degree law~\cite{BBCR03,HiBa08}. 
Actually, real company control and investment networks in general
adhere to a scale-free topology for reasons of portfolio differentiation~\cite{GBCStr}.

\longpaper{
Scale-free networks are characterized by a set of parameters (which we will discuss in detail),
that basically allow to control the mentioned distribution and, thus, give shape to each specific network~\cite{BBCR03}. They are: the expected number of nodes \(n\); the probability \(\alpha\) of adding a new node connected to an existing node chosen randomly according to the in-degree distribution; 
the probability \(\beta\) of adding an edge between two existing nodes; the probability \(\gamma\) of adding a new node connected to an existing node chosen randomly according to the out-degree distribution.}
From the available real-world cases, we learnt the values for the parameters of our scale-free networks
with specialized machine-learning techniques~\cite{Agar11}
\longpaper{, and obtained
\(\alpha = 0.71\), \(\beta=0.09\), \(\gamma=0.2\). We then used such values to generate random
scale-free networks with following the Erd\H{o}s-R\'enyi model, which we used to
build 7 artificial graphs, from 10 to 1 million companies.} We then considered the following scenarios.
\emph{AllRand}, where we 
ask for all the control relationships; \emph{QueryRand}, where we ask for 10 separate
pairs of companies and average the query times. 
We tested on 7 artificial graphs from 10 to 1M companies.

{\bf Results}. The results are reported in Figure~\ref{fig:experiments}(e-f). We observe that
the growth is much slower than in \emph{PSC} where we also explored company graphs.
Indeed, the characteristics of the graph of financial companies seem to make it simpler to explore
and the reasoning terminates in less than 10 seconds for \emph{AllReal} and less than
4 in \emph{QueryReal} for up to 50K companies. We observe that our synthetic graphs 
are a very good approximation of the real ones,
as the behavior of \emph{AllRand} and \emph{QueryRand} almost overlaps with the respective
real cases up to 50K companies. Results for very large graphs (100K and 1M companies) are
extremely promising, as the computation terminates in about 20 seconds for both 
\emph{AllRand}  and \emph{QueryRand}. We also point out that unlike \emph{SpecStrongLinks},
here restricting the reasoning to specific pairs of companies 
does not make an extreme difference, which we motivate
with the different graph topology: shorter chains and many hub companies.

\subsection{Comparison with Chase-based Tools}
\label{sec:chasebench}

\noindent
In Section~\ref{sec:iBench} we have evaluated the Vadalog system
on a set of non-trivially warded rules and used such results to validate
the performance against a range of chased-based systems.
In this section,
we complement the contribution with a validation of our system for the cases
of sets of rules just ``warded by chance'', i.e., the ones where there is a prevalence
of harmless joins, without any propagation of labelled nulls. In other terms,
in this section we conduct a comparative evaluation  (still relying on {\sc ChaseBench}) 
of settings resembling \emph{SynthG} in
Section~\ref{sec:synth}, typical of pure Datalog and data exchange/data integration scenarios,
modeled, e.g., with s-t tgds.

{\bf Description of the scenarios}. \emph{Doctors, DoctorsFD}: it is a data integration task from the schema mapping literature~\cite{MePS14}, even non-recursive yet rather important as a plausible
real-world case. We used source instances of 
10K, 100K, 500K, 1M facts and ran 9 queries of which we report the average times. 

\emph{LUBM}: The Lehigh University Benchmark (LUBM)~\cite{GuPH05} is a widely used synthetic benchmark
describing the university domain.
It is provided with a rule generator parametric in 
the number of universities.
We used source instances of 90K, 1M, 12M, and 120M facts. 
We ran 14 queries and averaged the answering time.

{\bf Results}. In Figure~\ref{fig:experiments}(g-i) we report the results, 
comparing our reasoner with the two top-performing systems on the scenarios at hand according
to {\sc ChaseBench}, namely RDFox and {\sc LLunatic}. The Vadalog system showed a very
promising behavior also for these cases where our termination and recursion
control algorithm cannot be exploited because of the very high fragmentation of the
warded forest induced by the absence of warded joins. In particular, 
on \emph{DoctorsFD}, the Vadalog system outperforms both the systems, being
3.5 times faster than RDFox and 2.5 faster than {\sc LLunatic}. On \emph{Doctors}
and \emph{LUBM}, our reasoner is up to 2 times faster than {\sc Llunatic} and one time slower than RDFox. 
This is motivated by the fact that, unlike RDFox, we do not incorporate yet specific 
Datalog optimization techniques, such as magic sets~\cite{AbHV95}, which
will certainly boost performance in such generic cases.

\cbstart
\vspace*{-1em}
\new{\subsection{Benchmarking the Lifted Linear Forest}
\label{sec:homomorphism}}

\noindent
\new{
With the experimental evaluation we describe in this section, we verify that the adoption of
lifted linear forests provides visible advantage over 
pure isomorphism check on the entire set of generated facts 
(the ``trivial technique'' mentioned in Section~\ref{sec:harmless}).
}

\new{{\bf Description of the scenarios.} We implemented a termination strategy based on  exhaustive storage of the generated facts and pure isomorphism check, and plugged it in into the Vadalog system. 
The code has been carefully optimized with hash-based indexing of the facts to have 
constant-time retrieval of previously generated homomorphic facts. 
We tested the Vadalog system comparing 
the performance of our approach (the full technique) with the trivial technique, using the
\emph{AllPSC} scenario described in Section~\ref{sec:dbpedia}.}

\new{{\bf Results.} As shown in Figure~\ref{fig:homomorphism},
for less than 100k persons, the trivial technique exhibits the same performance as
the full one, even being 1 or 2 seconds faster in some cases.
For more than 1M persons, the elapsed time for isomorphism check departs from the previous trend: the execution requires 64 seconds vs 52 seconds, observed with the linear forest. For 1.5M persons, 
the performance divergence grows even more and we observe 290 vs 86 seconds. Finally,
to have a clearer picture of the trend, we specifically extended the \emph{AllPSC} scenario 
and also ran it with 2M persons (from synthetic data) and obtained 549 vs 125 seconds.}
\new{The experiment shows that the trivial technique does not scale even when a very efficient 
data structure is adopted. 
As a matter of fact, isomorphism check requires storing all the generated facts; 
for small input instances, a very efficient indexing technique pays off. 
After a ``sweet spot'' at 100k, where the two algorithms basically have the same performance, 
for larger input instances isomorphism check times grow by orders of magnitude as the indexing structure eventually becomes inefficient (e.g., in the case of a hash structure, for example, for the soar in the number of hash conflicts) or even overflows the available heap space, requiring costly off-heap and to-disk swappings. As a further confirmation of these results, we ran the same test again with
all our cases, which confirmed the trend: for example for the \emph{AllRand} scenario
presented in Section~\ref{sec:industrial_validation}, elapsed time grows to 62 seconds for 1M persons.}

\new{
\begin{figure} [t!]
\centering
    \includegraphics[scale=0.25]{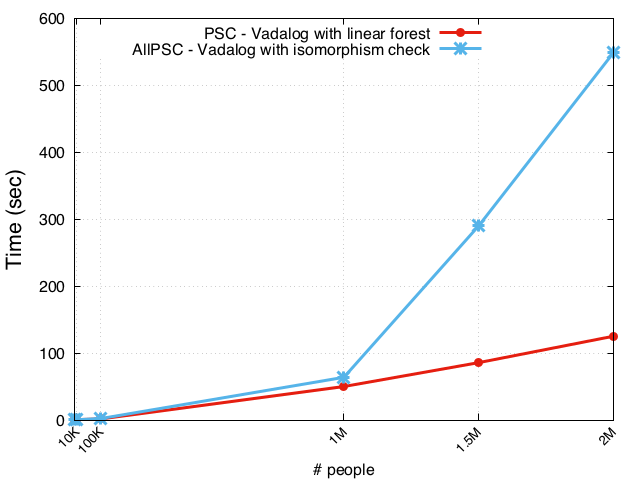}
\caption{DBpedia PSC with and w/o isomorphism check.}
\label{fig:homomorphism}
\end{figure}
}

\vspace*{-2em}
\new{
\subsection{Scaling Rules and Tuples}
\label{sec:dbsize}
}

\noindent
\new{In order to fully characterize the scalability of the Vadalog system, 
we evaluated reasoning performance along further dimensions: size of the source database,
number of rules, number of atoms in the rules, arity of the predicates.}

\new{{\bf Description of the scenarios.}
All our scenarios are variations of \emph{SynthB}, introduced in Section~\ref{sec:synth}.}

\new{\emph{DbSize:} 
We ran \emph{SynthB} with an initial database of 10k, 50k, 100k and 500k source instances. 
We generated input facts adopting a uniform 
distribution of values to induce an average join rate.}

\new{\emph{Rule\#:} 
We considered variants of \emph{SynthB} with 100, 200, 500, 1k rules respectively. We started from the basic version with 100 rules, with 50 input predicates and 10 output predicates. Then, we built
larger scenarios by composing blocks of rules. Each block is a copy of the basic \emph{SynthB} 
(with 100 rules), with rules appropriately renamed and linked to the respective input and output predicates. 
In this way, each set of rules has the same ``reasoning complexity'' 
as the basic case because independent 
blocks do not interact with each other, 
yet isolate the effect of scaling up the number of rules.}

\new{\emph{Atom\#:}
We considered variants of \emph{SynthB} having join rules with an average of 2, 4, 8 and 16 atoms. 
We started from the basic version with binary joins (bodies with 2 atoms) and scaled up, 
adding atoms to the bodies. 
Atoms have been added in such a way that the number of output facts is preserved and 
are proportionally distributed among the 27 recursive join rules and the 
63 non-recursive rules not to overly alter the computational 
complexity of the original rule set and actually test the effect of ``larger'' rules. 
The 10 linear rules have been left unmodified.}

\new{\emph{Arity:}
We constructed variants of \emph{SynthB} with atoms having 
an average arity of 3, 6, 12, 24. In order to test the effect of ``larger'' tuples in isolation, 
we increased the arity by adding variables that do not produce new 
interactions between atoms, so the proportion of 
harmless and harmful joins is also maintained.}

\new{{\bf Results.}
\emph{DbSize} (Figure~\ref{fig:scaling}(a)) confirms the polynomial behavior of the
warded fragment and good absolute elapsed times. The trend
is clearly polynomial (concretely, slightly sublinear): for 10k the obtained elapsed time is 4 seconds; 
it grows to 9 seconds for 50k, 13 seconds for 100k and 51 seconds for 500k.
Let us come to \emph{Rule\#} (Figure~\ref{fig:scaling}(b)). For 100 rules, we have  
the usual elapsed time for \emph{SynthB} (9 seconds), stable for 200 rules. 
We obtained 52 seconds for 500 rules and 101 seconds for 1k rules. 
The results confirm that the system is capable of handling an increasing 
number of rules and the performance is not affected by this. 
In fact, there is a linear growth of the elapsed times when 
multiple independent reasoning tasks are added as an effect of more rules.
In other terms, the effect of scaling up the number of rules is negligible
and only reflects in a slight increase in the time required for 
harmful-harmful simplification, construction of the execution plan and so on.
For \emph{Atom\#} (Figure~\ref{fig:scaling}(c)), starting from the base case with 2 body atoms (9 seconds), we obtained 10 seconds for 4 atoms, 18 seconds for 8 atoms and 27 seconds for 16 atoms. 
The execution optimizer implements multi-joins as a cascade of single-joins, 
and thus adding more atoms corresponds to lengthening the processing pipeline. 
As the size of the input instance does not vary, indexes allow to limit the impact of multiple joins, 
with a growth of execution time clearly polynomial (concretely, slightly sublinear).
Finally, \emph{Arity} (Figure~\ref{fig:scaling}(d)) shows that the size of facts (source and intermediate results) 
is almost irrelevant with respect to performance;
starting from the base case with the usual elapsed time of 9 seconds, we obtain 9 seconds for arity 6, 10 seconds for arity 12 and 12 seconds for arity 24. The 1-2 seconds difference is due 
to the natural increase in the data size and the internal hash-based caching 
structures are not affected by a reasonable growth in tuple size.}

\new{
\begin{figure} [t!]
\centering
    \includegraphics[scale=0.35]{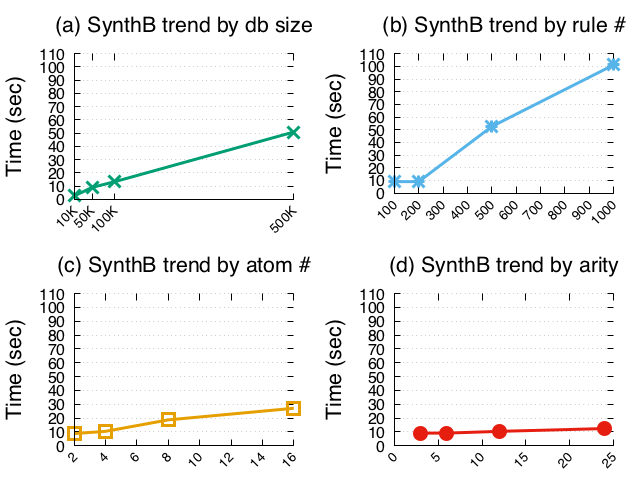}
\caption{Reasoning time scaling db size, rule size and arity.}
\label{fig:scaling}
\end{figure}
}
\cbend

\section{Related Work and Comparison}
\label{sec:relwork}

\noindent
A wide range of systems can be seen as related work to the Vadalog system. Let us first focus on the reasoners that share much (but not all) of the expressive power of our system, \cbstart \new{namely those that support (recursive) Datalog with existential quantification. Such systems have typically been applied in the areas} \cbend  of data exchange, integration, cleaning and query answering.\old{, there are plenty of systems.} \new{{\sc ChaseBench}~\cite{BKMM17} provides both an excellent collection of benchmarks  \cbstart  (most of which were presented in the preceding section) as well as a curated selection of systems in the area:} \cbend Graal~\cite{BLMR15}, {\sc LLunatic}~\cite{GMPS14}, RDFox \cite{MNPH14}, and PDQ~\cite{BeLT15}, as well as DEMo~\cite{PiSa09}, Pegasus~\cite{Meier14}, E \cite{Schulz13}, and ChaseFUN~\cite{BonifatiIL16}. \old{Many of them provide very good implementations of the chase~\cite{BKMM17}. However, as they are mostly conceived as traditional reasoners, their architecture
does not have a specific emphasis on scalability and so the languages they adopt.
Also, they often lack specific support for features required in important business scenarios and tend to focus on the strength and expressive power of the logic approach rather than the architecture required for enterprise settings. Datalog-based reasoners, such as DLV$^E$~\cite{LMTV12}, share many of these characteristics~\cite{LPFE06}.}

\cbstart
\new{
\textbf{PDQ} is a system based on query rewriting. More particularly, it reformulates a given query as query containment check under tgds and uses an underlying implementation of the restricted chase (i.e., chase with full homomorphism check) on top of PostgreSQL. The combination of restricted chase and rewriting into target systems’ language was shown not to be effective performance-wise, as systems fail to efficiently interleave homomorphism checks and updates~\cite{BKMM17}. PDQ is outperformed by the Vadalog system by three orders of magnitude in the tested scenarios.
Actually, PDQ's approach is effective in cases when the query reformulation drastically reduces the size of the input instance; however, this rarely happens in well-designed reasoning tasks (e.g., iBench) where the rules are compact and non-redundant. Furthermore, as we show in the industrial validation in Section 5.4, RDBMSs perform badly in the presence of recursion and the experimental results we present for PostgreSQL are consistent with the ones of PDQ. \textbf{LLunatic} is a system motivated by data cleaning settings, implementing several variants of the chase, including restricted chase with marked nulls. It is implemented on top of PostgreSQL, thus shares many of the characteristics in this regard as PDQ.
}

\new{
\textbf{Graal} is a toolkit for computing certain query answering under constraints. It adopts a saturation algorithm, which is basically a variant of the non-na\"ive restricted chase, where a fact is generated only if it passes the homomorphism check and the applicable rules are selected on the basis of a dependency graph, which avoids an exhaustive breadth-first search of applicable rules. The approach is based on forward chaining. It also embodies very smart query rewriting algorithms, which carefully take into consideration and implement specific logic fragments. Although Graal recognizes specific logic fragments (e.g., linear rules), such awareness is used only for query optimization purposes (e.g., rewriting), whereas the Vadalog system exploits the fragment structure to avoid homomorphism check, and, as a consequence, the need to store a fresh memory copy of all the generated facts. Conversely, Graal performs homomorphism check, implementing it against several different back-ends, with both the performance drawbacks of full homomorphism check 
(shown in Section~\ref{sec:harmless}) and those of the restricted chase-based implementations~\cite{BKMM17}.}

\new{
\textbf{DLV} is a mature Datalog system, which supports many different features. It covers disjunctive Datalog under the answer sets semantics (with two forms of negation). In the version considered by \cite{BKMM17}, existentials have been simulated by Skolemization. DLV implements most advanced Datalog optimization techniques such as magic sets, back-jumping, smart pruning and parallel processing. DLV performs better than PDQ and Graal as it adopts unrestricted chase and a pure in-memory processing. The main limitation of the in-memory version of the system (the only one supporting tgds), is a too large memory footprint due to grounding.
This becomes prohibitive for large input sizes and is the principle difference between DLV and the Vadalog system (which adopts a form of \emph{lifted} inference, i.e., does not require grounding).
}

\new{
\textbf{RDFox} is a high-performance in-memory Datalog engine, implementing both restricted and unrestricted chase and fully supporting existentials. The Vadalog system is three times faster than RDFox on iBench.
A principle difference between RDFox and Vadalog is that RDFox considers all rule instances, while Vadalog considers -- utilizing the lifted linear forest -- a reduced number of rule instances. This is particularly relevant if there are a high number of input tuples that are structurally similar (i.e., follow the same \textit{pattern} as described in Section 3).
}

\new{
Many more systems exist than those discussed in detail. In addition to the ones discussed above, \cite{BKMM17}  considers DEMo, Pegasus, E and ChaseFUN. Of these, ChaseFUN does not support recursive rules, and DEMo performs poorly on all the experiments \cite{BKMM17}. E is a constraint solver requiring processing one tuple at a time, and Pegasus ``fared badly on all tests'' due to it being based on query reformulation \cite{BKMM17}. However note that none of these systems was designed or optimized for the considered scenarios.
A further interesting system is LogicBlox~\cite{ACGK15}. Unfortunately, it was not available for benchmarks. 
Looking further, there are some similarities to the OnTop system~\cite{CalvaneseCKKLRR17}, the OBDA approach in general, as well as graph databases. Moreover, our basic architecture shares similarities with those of a huge variety of state-of-the-art data processing systems. Yet, as can be expected (but is also shown by our experiments -- in particular those in Section 5.3), systems that have no strong optimization for recursion or existential quantification are at a significant disadvantage performance-wise.
}

\new{
Some general themes emerge from the specific systems discussed above as well as those sharing similar properties: (a) \textbf{Restricted vs unrestricted chase}. Back-end based 
implementations of restricted chase are problematic. Let us, for example,
consider a small set of rules inspired by \emph{ONT-256} scenario,
introduced in Section~\ref{sec:iBench}.}
\new{\begin{example}
\label{ex:restricted_chase}
$D = \{{\rm Whistle}(1,1,2,3), {\rm Young}(1)\}$\\
$1: {\rm  Whistle}(a,a,b,c)  \rightarrow {\rm Whistle}(b,b,a,c)$\\
$2: {\rm  Whistle}(a,a,b,c)  \rightarrow \exists h~{\rm Cow}(a,b,\hat h)$\\
$3: {\rm Cow}(a,b,\hat h), {\rm Young}(a) \rightarrow {\rm Cow}(b,a,\hat h).$
\end{example}}

\new{In systems based on RDBMS backend, 
homomorphism check is implemented in SQL, so for rule 2, we would have:}

\begin{lstlisting}[
           language=SQL,
           showspaces=false,
           basicstyle=\ttfamily,
           numberstyle=\tiny,
           commentstyle=\color{green}
        ]
SELECT * FROM Whistle W WHERE NOT EXISTS 
  (SELECT * FROM Cow C WHERE C.a = W.a 
      and C.b = W.b) and W.a = W.a1
\end{lstlisting}
\noindent\new{If the query returns an empty result, the check is considered successful. 
Such an SQL query would have to be executed before each chase step,
and running it just once for the entire predicate ${\rm Whistle}$ would be insufficient.
To highlight this aspect, let us suppose rule~1 is first applied to 
$D$, generating ${\rm Whistle}(2,2,1,3)$. Now,
an update in ${\rm Cow}$ by rule~2, generating ${\rm Cow}(1,2,h_1)$,
may trigger rule~3, hence generating ${\rm Cow}(2,1,h_1)$; this 
invalidates the result of the previous check for rule~2, 
which would generate ${\rm Cow}(2,1,h_2)$ from ${\rm Whistle}(1,1,2,3)$.
Although many optimizations are possible, systems implementing backend-based 
chase (e.g., Graal, PDQ and to some extent {\sc LLunatic}) pay such query overhead, unlike 
systems that do not translate homomorphism checks into target system queries. In particular,
in this case the Vadalog system would recognize isomorphic copies
of ${\rm Cow}(1,2,h)$ in the same component of the warded
forest and thus inhibit rule 3 from firing. This would be done by
recognizing the initial pattern \({\rm Whistle}(c_1,c_1,c_2,c_3)\) in the
lifted-linear forest and cutting at the stop provenance.
(b) \textbf{In-memory}.
The restriction to in-memory processing is successful, yet it causes a too large memory footprint, in particular in combination with grounding, and fails for Big Data. In contrast, the Vadalog system exploits the specificities of Warded Datalog$^\pm$ to implement 
a restricted chase that optimizes the search for isomorphic facts by
restricting to the same component of the warded forest and 
adopting good guide structures, that also limit memory footprint.
For instance, a potentially very large set of components in the lifted-linear 
forest, pattern-isomorphic to the one in Example~\ref{ex:restricted_chase}, 
can be represented by just a single pattern component with five facts.
 }
\old{
Among the Datalog-based reasoners that share the expressive power of the {\sc Vadalog} system, of particular note are RDFox \cite{MNPH14} and LogicBlox~\cite{ACGK15}. These two systems are very good reasoners, perform very well in benchmarks~\cite{BKMM17} (RDFox) and cover many features (LogicBlox).
Unfortunately, LogicBlox was not available for benchmarking. 
Yet, we follow a completely different approach in how we achieve good performance: we give real value to the adopted logic fragment and propose data structures that exploit the specific underpinnings of the language in the context of a modern data management architecture. In particular, the {\sc Vadalog} system is the first implementation of the Warded Datalog$^\pm$ language.}
\cbend

\smallskip
\section{Conclusion}
\label{sec:conclusions}

\noindent
In this paper, we introduced the Vadalog system, the first implementation of Warded Datalog$^\pm$. At the core of it, the algorithm exploits the key theoretical underpinnings of the underlying language through novel recursion control techniques. The architecture and system exhibited competitive performance in both real-world and synthetic benchmarks.

As future work, we are working on adding further standard data\-base optimization techniques to the system, such as more sophisticated cost-based query plan optimization. We also are constantly expanding our range of supported data sources so as to encompass more and more data stores, big data platforms and queryable APIs.

More speficially, we plan to work on introducing consistent query answering \cite{pods/ArenasBC99,icdt/ArmingPS16} to the system as well as work on view updates \cite{pods/BunemanKT02,amw/GuagliardoPS13}. We also intend to apply our system to big data settings in the area of computational social choice, such as huge elections \cite{aaai/CsarLPS17}.

\medskip
\noindent
\textbf{Acknowledgements}. The work on this paper was supported by EPSRC programme grant EP/M025268/1 VADA.
The Vadalog system as presented here is IP of the University of Oxford.

\balance

\bibliographystyle{abbrv}
\bibliography{b}

\end{document}